\documentclass[prc,aps,superscriptaddress,nofootinbib,twocolumn]{revtex4}
\usepackage{amssymb}
\usepackage[dvips]{graphicx}
\usepackage[english]{babel}
\makeatletter\AtBeginDocument{\let\@elt\relax}\makeatother

\usepackage{indentfirst}
\usepackage{amsxtra}
\usepackage{amsmath}
\usepackage{supertabular}
\usepackage{multirow}
\usepackage[mathcal]{eucal}
\usepackage[usenames]{color}
\usepackage{ulem}

\begin{document}
\title {\textbf{Finite-temperature infinite matter with effective-field-theory-inspired energy-density functionals}}
\author{Stefano Burrello}
\affiliation{Technische Universit\"{a}t Darmstadt, Institut f\"{u}r Kernphysik, Darmstadt, Germany}
\affiliation{Universit\'e Paris-Saclay, CNRS/IN2P3, IJCLab, 91405 Orsay, France}
\author{Marcella Grasso}
\affiliation{Universit\'e Paris-Saclay, CNRS/IN2P3, IJCLab, 91405 Orsay, France}

\begin{abstract}
 
Finite-temperature infinite matter is analyzed with the recently introduced 
effective-field-theory(EFT)-inspired YGLO (Yang-Grasso-Lacroix-Orsay) and ELYO (extended Lee-Yang, Orsay) functionals, which are designed to describe very low-density regimes in symmetric (YGLO) and in pure neutron (YGLO and ELYO) matter. The article deals with neutron matter 
and aims to verify whether 
the use of these functionals allows us to correctly incorporate finite-temperature effects. We compare our results for some relevant thermodynamical quantities 
with the corresponding ones computed with a chosen reference {\it{ab-initio}} model, namely the many-body-perturbation-theory scheme. We validate the reliability of both EFT-inspired functionals at least at rather low densities and not too high temperatures and we discuss the effects related to the effective mass.  We conclude that, at the present stage, the ELYO functional, having a higher neutron effective mass around saturation (closer to {\it{ab-initio}} values), allows us to describe finite-temperature properties more satisfactorily, in better agreement with {\it{ab-initio}} predictions up to higher densities and temperatures, compared to YGLO. 

\end{abstract}

\maketitle

\section{Introduction}
\label{intro}
Energy-density-functional (EDF) theories are extensively adopted to deal with the nuclear many-body problem. They allow for a unified treatment of static properties of nuclei across the whole nuclear chart and their time-dependent versions provide a reliable description of a large variety of dynamical phenomena \cite{bender,colo,schunck, nakatsukasa,stone}.
Starting from the pioneering work of Ref. \cite{negele}, EDFs have also been widely applied to treat the isospin-asymmetric nuclear systems located in the crust of neutron stars (NSs), giving insight into several properties of astrophysical interest \cite{fantina, potekhin, pearson1, pearson2, burrello15}. 

Nevertheless, it is well known that most nuclear EDFs are built on empirical ingredients. This represents a clear limitation and reduces their  predictive power far beyond the domain where the adjustment of their parameters is performed. In addition, 
differently from effective-field theories (EFTs), traditional EDFs are not developed following a power-counting scheme, which would allow for setting up an order-by-order hierarchy of the interaction contributions.   Incidentally, 
it is interesting to notice that a few steps towards the construction of a power counting in EDF were recently reported in Refs. \cite{yangpc,burrello}, based on a scheme inspired by the Dyson many-body expansion. 

In recent years, several attempts have been made to render EDFs less empirical, trying to directly link them to microscopic ingredients so to reduce their intrinsic uncertainties (see for instance Refs. \cite{fur1,fur2,grasso2019}). For example, a special class of functionals inspired by EFTs and benchmarked on {\it{ab-initio}} predictions have been designed \cite{yglo2016,elyo2017,drop1,drop2,yglonuclei}. These functionals have been called YGLO (Yang-Grasso-Lacroix-Orsay) and ELYO (extended Lee-Yang, Orsay). Their peculiarity is to properly describe infinite matter both at densities close to the saturation point of symmetric matter, of interest for nuclear phenomenology, and at very low densities, where the dilute gas is described by a Lee-Yang (LY) expansion \cite{LY1,LY2,LY3,LY4,LY5,LY6}. The YGLO functional has been recently applied also to finite nuclei \cite{yglonuclei}. 

The aim of this work is to employ these EFT-inspired EDFs, originally introduced at zero temperature, for addressing finite-temperature properties of infinite nuclear matter (NM) and for describing its thermodynamical behavior \cite{arianna}. An accurate description of the NM behavior, over a wide range of densities and temperatures, is indeed mandatory to provide a global and multi-purposes formulation of the equation of state (EOS), to be employed in particular for modeling compact stellar objects \cite{oertel}. 
Indeed, except for the case of cold (catalyzed) NSs, for which the zero-temperature approximation is used,
finite-temperature NM EOSs  have been shown to be important for example in proto-NSs and core-collapse supernovae computations (e.g., in Refs. 
\cite{schneider,yasin}). The proper treatment of finite-temperature EOSs is 
crucial also
in hydrodynamical calculations for NS mergers. For the latter, the most widely used finite-temperature EOSs are based on Skyrme-like models or on relativistic mean-field approaches (see Refs. \cite{oertel, burgio} for a list of multi-purposes EOSs) and describe the thermal part as an ideal fluid, although this approximation has revealed  itself to be quite crude to mimic thermal properties of the EOS \cite{bauswein, constantinou, arianna2}.
Reference \cite{bauswein} showed for example the importance of a consistent treatment of thermal effects in the EOS and the implications in particular for merger simulations, for which the approximation of parametrizing the temperature dependence of the EOS through a constant thermal index (between 1 and 2)  \cite{shibata1,shibata2,shibata3,kiuchi,janka} reveals itself to be quite poor.

For this reason, different approaches have been recently employed to extend to finite temperatures the EOS of cold and homogeneous nuclear matter \cite{du, raithel}.

In the present work, with the aim to move along this direction and validate the reliability of the recently introduced EFT-inspired functionals at finite temperatures, we  focus on pure neutron matter (PNM) and we compare our results with selected microscopic {\it{ab-initio}} predictions. 
Among the available finite-temperature microscopic calculations for matter \cite{FP,zuo,tolos,fiorilla,carbone1,welle1,welle2,welle3,carbone2,rios,lu,keller}, we have adopted as reference computations (to which compare our results) the recent many-body-perturbation-theory (MBPT) predictions for PNM 
of Ref. \cite{keller}, obtained with two- and three-nucleon interactions,  
constructed within chiral EFT up to next-to-next-to-next-to-leading order. 
The spread produced in the results of Ref. \cite{keller} owing to the different choices of 
chiral interactions 
covers most of the uncertainties existing in 
other available 
ab-initio calculations at high densities and temperatures \cite{FP,zuo,tolos,fiorilla,carbone1,welle1,welle2,welle3,carbone2,rios,lu,keller}.  We notice moreover that, in the low-density and temperature regime we are mostly focusing on in the present work, these uncertainties are 
strongly reduced and fully accounted for within the bands provided in Ref. \cite{keller}, whose MBPT results were 
chosen for this reason as reference computations in our work.
In particular, we compare 
several thermodynamical quantities such as 
the entropy per nucleon, the free energy per nucleon, the pressure, and the thermal index and we discuss the role played by the effective mass. 

The article is organized as follows. Section \ref{functional} reminds the expressions for the functionals and the effective masses. We present in Sec. \ref{thermo} the formulas for the entropy, the free energy, the pressure, the thermal index, and the chemical potentials. We illustrate in Sec.  \ref{resu} the  predictions  obtained for PNM  and discuss the comparison with the corresponding reference {\it{ab-initio}} results.  Conclusions are drawn in Sec. \ref{conclu}. 

\section{Functionals and effective masses}
\label{functional}
\subsection{Skyrme-like EDFs}

The EDF generated at leading order by employing a Skyrme-like effective interaction \cite{sk1,sk2,vau} to describe spin-saturated NM may be written as
\begin{equation}
\label{eq:EDF} 
\mathcal{E} =  \mathcal{K} + \mathcal{E}_{Sk} = \mathcal{K} + \mathcal{E}_0 + \mathcal{E}_3 + \mathcal{E}_{\textup{eff}},
\end{equation}
where $\mathcal{K} = \frac{\hbar^2}{2m} \tau$ is the kinetic term, $\tau$ being the isoscalar kinetic energy density, $\tau = \tau_n + \tau_p$. 
We denote by $m = 939$ MeV the bare nucleon mass. Indices $n$ and $p$ stand for neutrons and protons, respectively. The potential part $\mathcal{E}_{Sk}$, sum of three contributions, may be expressed 
in terms of isoscalar ($\rho = \rho_n + \rho_p$ and $\tau$) and isovector ($\rho_3 = \rho_n - \rho_p$ and $\tau_3 = \tau_n - \tau_p$) matter and kinetic energy densities as
\begin{eqnarray}
\label{eq:skyrme_parameters}
\nonumber
\mathcal{E}_{Sk} &=& C_0 \rho^2 + D_0 \rho_3^2 + \left (C_3 \rho^2 + D_3 \rho_3^2 \right ) \rho^\alpha \\ 
&+& C_{\textup{eff}} \rho \tau + D_{\textup{eff}} \rho_3 \tau_3. 
\end{eqnarray}
The coefficients $C$ and $D$ are expressed in terms of the traditional Skyrme parameters ($\alpha$, ($t_i, x_i$), $i=0,\dots,3$), as follows:
\begin{eqnarray}
\label{cd_skyrme}
C_0&=& \dfrac{3}{8} t_0; \; \; C_3= \dfrac{1}{16} t_3; \nonumber \\ 
C_{\textup{eff}}&=& \dfrac{1}{16} \left [ 3 t_1 + t_2  \left (5 + 4 x_2 \right ) \right ]; \; \;  
\\
D_0 &=& - \frac{1}{8} t_0 \left (1 + 2 x_0 \right );  \; \; 
D_3 = - \frac{1}{48} t_3 \left (1+ 2 x_3 \right ); \nonumber   \\ 
D_{\textup{eff}} &=& - \frac{1}{16} \left [  t_1 \left (1 + 2 x_1 \right ) - t_2 \left ( 1+ 2 x_2 \right ) \right ].   \nonumber 
\end{eqnarray}
The effective masses may be deduced from the momentum-dependent part of the functional \cite{zheng2}. The effective mass for each nuclear species $q$ is defined as
\begin{equation}
\frac{\hbar^2}{2 m_q^*} = \frac{\partial \mathcal{E}}{\partial \tau_q}. \qquad \textup{q = p, n}
\end{equation}
Using Eqs.~\eqref{eq:EDF} and~\eqref{eq:skyrme_parameters}, we obtain
\begin{equation}
\label{mstar}
m_q^* = \frac{m}{1 + \frac{2 m}{\hbar^2}(C_{\textup{eff}}\rho \pm  D_{\textup{eff}}\rho_3)}, \qquad \textup{q = p, n}
\end{equation}
where the $+$ ($-$) sign holds for neutrons (protons).

Whereas infinite matter at zero temperature is not sensitive to the value of the effective mass, predictions for finite nuclei (see for example Ref. \cite{yglonuclei} for the YGLO case) and for finite-temperature matter are affected by this quantity.

\subsection{YGLO case}
All the details about the YGLO functional are published in Refs. \cite{yglo2016,drop1}.
The potential part of the functional for symmetric nuclear matter (SNM) ($i = s$) and PNM ($i = n$) is given by
\begin{equation}
\label{eq:yglo}
\mathcal{E}_{Y} = Y_i [\rho] \rho^2 + D_i \rho^{8/3} + F_i \rho^{(\alpha + 2)},
\end{equation}
where $Y_i [\rho]$ is the so-called resummed term \cite{yglo2016} which has the following analytical form:
\begin{equation}
Y_i [\rho] = \frac{B_i}{1 - R_i \rho^{1/3} + C_i \rho^{2/3}}.
\end{equation}
The expressions for $B_i$ and $R_i$ are
\begin{align}
\label{eq:bi_ri}
B_i &= \frac{2 \pi \hbar^2}{m} \frac{\nu_i - 1}{\nu_i} a_i, \nonumber \\ 
R_i &= \frac{6}{35 \pi} \left ( \frac{6 \pi^2}{\nu_i} \right )^{1/3} \left ( 11 - 2 \ln 2 \right ) a_i,  
\end{align}
where $\nu_i =$ 2 (4) is the degeneracy factor for PNM (SNM). $B_i$ and $R_i$ 
are fully constrained by the value of the $s$-wave scattering length, 
$a_i =$ -18.9 (-20.0) fm for PNM (SNM) (see Ref. \cite{yglo2016} for details). The values of the other parameters $C_i$, $D_i$, and $F_i$ may be found in Ref. \cite{yglo2016}, where the procedure of their adjustment on {\it{ab-initio}} predictions is also detailed. Two YGLO parametrizations were proposed in Ref. \cite{yglo2016}, YGLO(FP) and YGLO(Akmal). For both of them, the EOS of SNM follows the EOS of Ref. \cite{FP} beyond the low-density regime, driven by the LY expansion. On the other side, the EOS of PNM is adjusted on Quantum Monte Carlo (QMC) results \cite{gezerlis} between the very low-density regime and densities close to 0.005 fm$^{-3}$.  At densities larger than $\approx$ 0.005 fm$^{-3}$,
the YFLO(FP) (YGLO(Akmal)) PNM EOS is adjusted  on the PNM EOS of Ref. \cite{FP} (\cite{akmal}). 

Special attention deserves the $D_i$ term, since such a power of $\rho$ may originate from a momentum-dependent term $\mathcal{E}_{\textup{eff,Y}}$, from an extra density-dependent term $\mathcal{E'}_{3,Y}$ (with $\alpha'_Y$ = 2/3) or from any combination of both. 
To remove this ambiguity, a new coefficient was introduced in Ref. \cite{drop1}, called splitting parameter and denoted by $W$, which determines the distribution of $D_i$ into a momentum-dependent term and a density-dependent one, such that:
\begin{equation}
\mathcal{E}_{\textup{eff,Y}} = W D_i \rho^{8/3}, \qquad \mathcal{E'}_{3, Y} = (1 - W) D_i \rho^{8/3}.
\end{equation}
A splitting parameter different from zero generates an effective mass different from the bare mass, without modifying the EOS.  A mapping with Skyrme-type parameters may be carried out and the following relations may be derived for the coefficients of the density-dependent contribution $\mathcal{E'}_{3, Y}$:  
\begin{equation}
\label{ttt}
t'_3 = 16(1 - W) D_s , \qquad t'_3 ( 1 - x'_3) = 24(1 - W) D_n,
\end{equation}
in analogy with the other density-dependent term $\mathcal{E}_{3, Y}$ (depending on $F_i$ in Eq. (\ref{eq:yglo})), whose coefficients may be written as 
\begin{equation}
\label{t33}
t_3 = 16 F_s , \qquad t_3 ( 1 - x_3) = 24 F_n.
\end{equation}
For the momentum-dependent contribution $\mathcal{E}_{\textup{eff,Y}}$, the strategy followed in Ref. \cite{drop1} consisted in retaining only the $s$-wave part, that is $t_2 = x_2 = 0$. For the YGLO(FP) case, the adjustment on microscopic results available for neutron drops led to $W = -0.0840$ 
in Ref. \cite{drop1} and to $W = -0.0764$ in Ref. \cite{yglonuclei}. The two sets have only slightly different values for the effective masses. As an illustration, we use the YGLO(FP) parametrization of Ref. 
\cite{yglonuclei} throughout this work.  

The following relations hold: 
\begin{eqnarray}
\label{md}
t_1 = \frac{80}{9} W \left ( \frac{3 \pi^2}{2} \right )^{-2/3} D_s, \\
\nonumber
 t_1 (1 - x_1) = \frac{40}{3} W (3 \pi^2)^{-2/3} D_n.
\end{eqnarray}
The YGLO effective masses may be computed using Eq. (\ref{mstar}), where $C_{eff}$ and $D_{eff}$ are related to $t_1$ and $x_1$ ($t_2=x_2=0$) by Eqs. (\ref{cd_skyrme}) and $t_1$ and $x_1$ are related to $W$, $D_s$, and $D_n$ by Eqs. (\ref{md}).  

\begin{figure}[h]
\centering
\includegraphics[width=0.45\textwidth]{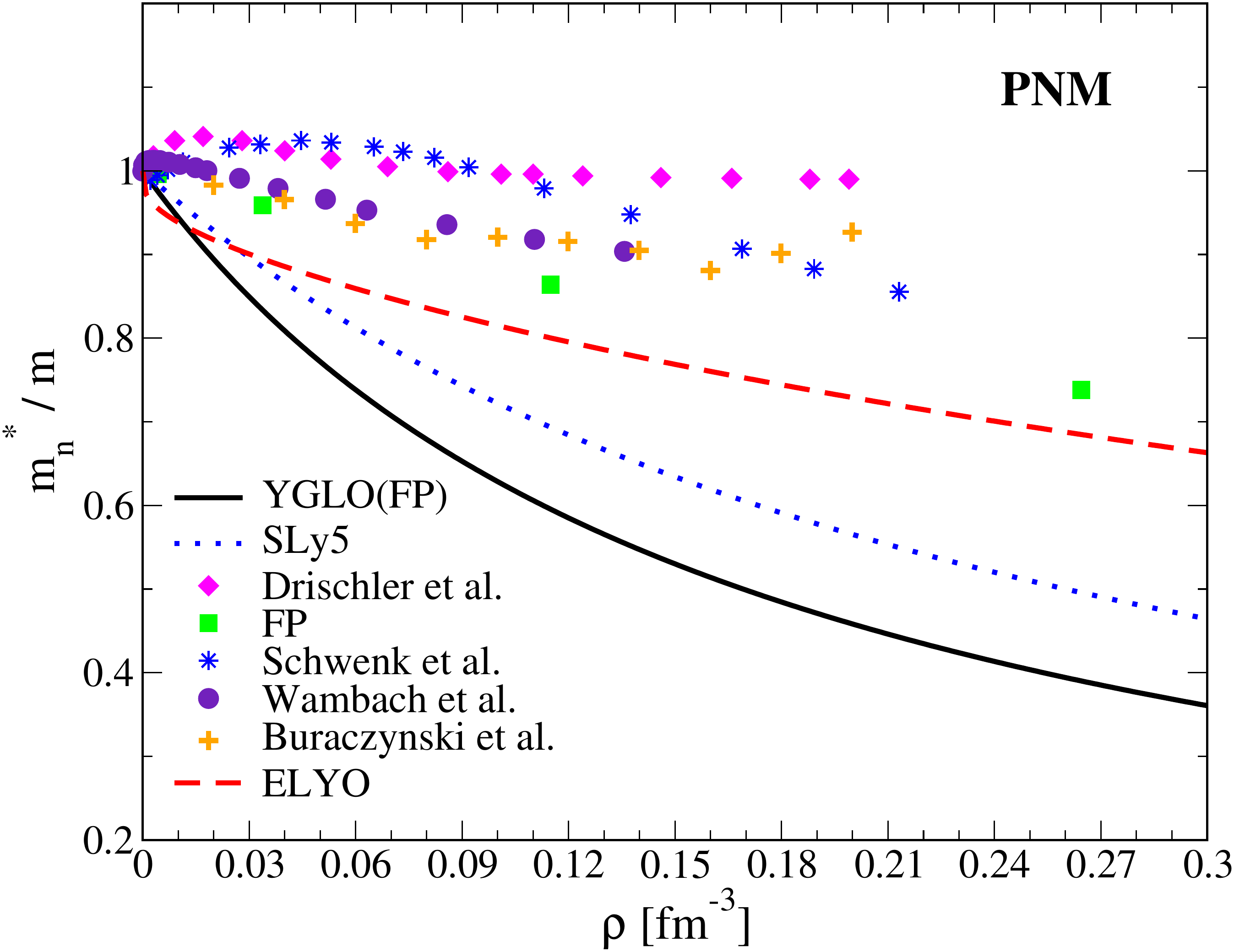}
\caption{Neutron effective mass as a function of the density, obtained with the YGLO(FP) functional of Ref. 
\cite{yglonuclei}, the ELYO functional of Ref. \cite{drop2}, and the Skyrme SLy5 functional. We also display the neutron effective masses extracted from Refs. \cite{FP,dri,sch,wam} 
and from recent QMC results with chiral interactions, including 
three-body contributions (Refs. \cite{bura1, bura2})}. 
\label{fig:effective_mass}
\end{figure}

The impact of the effective mass on the results for PNM will be analyzed in Sec. \ref{resu}. The effective mass in this case is the neutron effective mass. 
We remind here that, in the YGLO case (with the parameter sets of both Ref. \cite{drop1} and Ref. \cite{yglonuclei}), the neutron effective mass is very small compared to 
 microscopic {\it{ab-initio}} predictions and to many Skyrme values.  This may be seen in Fig. \ref{fig:effective_mass}, where, together with 
the YGLO(FP) 
\cite{yglonuclei} effective mass, the corresponding Skyrme SLy5 \cite{cha} values are also plotted, as an illustration. Also some available {\it{ab-initio}} predictions are displayed \cite{FP, dri, sch, wam, bura1, bura2}. 

It was suggested in Ref. \cite{yglonuclei} that 
possible directions for increasing the YGLO effective mass 
(still benchmarking on ab-initio results) 
may be foreseen by enlarging this functional in order to include also $p$-wave contributions 
and by carrying out once again the adjustment procedure on neutron-drop energies. 
This would of course require some dedicated work which is planned for future studies. Here, in order to visualize as an illustration the effect that a higher YGLO effective mass might have, we modify the functional by changing by hand the value of its splitting parameter, without readjusting the other parameters. In this way, the EOS of matter is not affected but we are aware that the predictions for neutron-drop energies are deteriorated compared to those obtained with the original set of parameters. 

\subsection{ELYO case}

The most recent version of the ELYO functional was designed, in Ref. \cite{drop2},  by writing the PNM EOS as the sum of the first terms of the LY expansion, up to the $(k_F^5)$ contributions, where $k_F$ is the Fermi momentum, $k_F=(3 \pi^2 \rho)^{1/3}$ in PNM. To ensure that this truncated expansion is able to describe equally well the PNM EOS, both at very low densities and at densities close to the saturation point, a density-dependent neutron-neutron $s$-wave scattering length $a_s (\rho)$ was adopted, naturally tuned by a low-density-type condition  $|a_s(k_F) k_F| = 1$ (assumed valid at all densities), with $a_s=$ - 18.9 fm close to zero density. 

A mapping may be performed with a Skyrme-type functional, Eq. (\ref{eq:skyrme_parameters}) and an adjustment on the equilibrium point of SNM may be carried out. Such a functional can be used to describe both SNM and PNM, by introducing the following relations: 
\begin{subequations}\label{condLYs}
\begin{align} 
&t_0(1-x_0)=\dfrac{4\pi\hbar^2}{m}a_s(\rho) ,\\
&t_3(1-x_3)=\dfrac{144\hbar^2}{35m} c_0 (11-2\ln 2)a_s^2(\rho), \\
& t_1(1-x_1)=W_1\dfrac{2\pi\hbar^2}{m} B_s (\rho), \\
& t_{3'}(1-x_{3'})=(1-W_1) \dfrac{36c_0^2\pi\hbar^2}{10m} B_s (\rho), \\
& t_2(1+x_2)=W_2\dfrac{4\pi\hbar^2}{m} a_p^3, \\
& t_{3''}(1-x_{3''})=(1-W_2)\dfrac{108c_0^2\pi\hbar^2}{5m} a_p^3 ,
\end{align}
\end{subequations}
where $a_p$ is the neutron-neutron $p$-wave scattering length, $c_0=(3\pi^2)^{1/3}$, and 
\begin{equation} \label{Brho}
B_s (\rho) \equiv \Bigl[ r_s a_s^2(\rho) +0.19\pi a_s^3(\rho)\Bigr].
\end{equation} 
We have introduced two splitting parameters in Eqs. (\ref{condLYs}), $W_1$ and $W_2$, which separate momentum-dependent and density-dependent contributions. The powers of the density in the three density-dependent terms are $\alpha=1/3$ and
 $\alpha'=\alpha''=2/3$.  
Using Eqs. (\ref{cd_skyrme}),
the parameters of Eqs. (\ref{condLYs}) define both the ELYO Skyrme-type functional, Eq. (\ref{eq:skyrme_parameters}), and the ELYO effective masses, Eq. (\ref{mstar}). 

The values of the parameters are reported in Ref. \cite{drop2} and the physical value of $a_p$ = 0.63 fm is used, corresponding to the AV4 interaction (see Ref. \cite{boulet}).

There is a typo in Table II of Ref. \cite{drop2}. The value of the splitting parameter $W_1$ is -0.176 (instead of -0.163). Differences in the predictions for neutron drops are however very small when the two values are used. We use in what follows the correct value $W_1=$ -0.176. 

With the ELYO  functional \cite{drop2}, effective masses are in general higher than with YGLO and, for the case of the neutron effective mass, in better agreement with {\it{ab-initio}} predictions, as was discussed in Ref. \cite{drop2}. This may be seen in Fig.  \ref{fig:effective_mass}.

\section{Thermodynamical properties}
\label{thermo}

In the EDF framework we may extract, in a given statistical ensemble, the corresponding thermodynamical potential. The Helmholtz free energy $F  = E - TS$, where $E$ and $S$ are, respectively, the total energy and the entropy of the system, is the thermodynamical potential which allows us to investigate 
a nuclear system with fixed numbers of neutrons $N$ 
at temperature $T$. We compute here  this quantity to analyze the properties of the EOS of PNM 
at different  densities and temperatures. 

For a nuclear system with $N$ neutrons 
at finite temperature $T$ in the thermodynamical limit, 
\begin{equation}
\frac{N}{V} \xrightarrow{N,V \to \infty} \rho
\end{equation}
the free energy comes out to be proportional to the volume $V$. 

\subsection{Zero temperature}
The free-energy density functional coincides with the EDF at zero temperature. As an illustration, by using the Skyrme-type functional (\ref{eq:skyrme_parameters}), the PNM energy per nucleon may be written in terms of the Skyrme parameters of Eq. \eqref{cd_skyrme} as 
\begin{eqnarray}
\label{eq:energy_nucleon}
\nonumber
\frac{E}{N} &=& \frac{E_{FG}}{N} + (C_0 + D_0)\rho + (C_3 + D_3)\rho^{\alpha+1} \\ 
&+& (C_{eff} + D_{eff})\tau,
\end{eqnarray}
where $\tau = \frac{3}{5}(3\pi^2\rho)^{2/3}\rho$ is merely a function of $\rho$ and the energy per nucleon of the free Fermi gas (FG) is given by 
\begin{equation}
\frac{E_{FG}}{N} = \frac{3}{5} \epsilon_F.
\end{equation}
In the case of the YGLO functional, 
using  Eq. (\ref{eq:energy_nucleon}) with $C_0=D_0=0$, 
one may write
\begin{equation}
\label{eq:energy_YGLO_PNM}
\frac{E_Y}{N} = \frac{E}{N} + Y_n \rho + (C'_{3,Y} + D'_{3,Y})\rho^{\alpha'_{Y}+1}, 
\end{equation}
where $C_3$ and $D_3$ in Eq. (\ref{eq:energy_nucleon}) are related to $t_3$ and $x_3$ by Eqs. (\ref{cd_skyrme}) and $t_3$ and $x_3$ are related to the YGLO parameters $F_s$ and $F_n$ by Eqs. (\ref{t33}); $C_{eff}$ and $D_{eff}$ in Eq. (\ref{eq:energy_nucleon}) are related to $t_1$ and $x_1$ by Eqs. (\ref{cd_skyrme}) (with $t_2=x_2=0$) and $t_1$ and $x_1$ are related to the YGLO parameters $D_s$, $D_n$, and $W$ by Eqs. (\ref{md}); finally, the coefficients  $C'_{3,Y}$ and $D'_{3,Y}$ of the additional density-dependent part in Eq. (\ref{eq:energy_YGLO_PNM}) are related to Skyrme-type parameters $t_3'$ and $x_3'$ by relations analogous to those in Eqs. (\ref{cd_skyrme}) and $t_3'$ and $x_3'$ are related to the YGLO parameters $D_s$, $D_n$, and $W$ by Eqs. (\ref{ttt}). We remind that $\alpha'_{Y}=$ 2/3. 

In the case of the ELYO functional, Eq. (\ref{eq:energy_nucleon}) holds, using the parameters of Eqs. (\ref{condLYs}) in Eqs. (\ref{cd_skyrme}). 

At zero temperature, the effective chemical potential is equal to the Fermi energy renormalized by the effective mass, 
$\tilde{\mu} = \epsilon_{F}^* = \frac{\hbar^2 k_F^2}{2 m_n^*}$. 

\subsection{Finite temperature}
At finite temperatures, to determine the effective chemical potential 
one has to solve the density equation, which generally is written as 
\begin{equation}
\label{eq:density}
\rho =\nu  \int \frac{d^3 p}{h^3} f(p)
\end{equation}
where $\nu = 2$ is the spin-isospin degeneracy factor and $f$ is the Fermi distribution function 
\begin{equation}
\label{fq}
f = \frac{1}{1 + \exp\left(\displaystyle{\frac{\epsilon - \tilde{\mu}}{T}}\right)}
\end{equation}
being $\epsilon = p^2/(2m^*_n)$. 
On the other hand, $\tau$, which is expressed as
\begin{equation}
\tau = \frac{\nu}{\hbar^2} \int \frac{d^3 p}{h^3} p^2 f(p) 
\end{equation}
turns out to be non-trivially dependent on $\rho$.

To evaluate the free energy per nucleon one has to compute the entropy per nucleon 
$S / N$, which reads
\begin{eqnarray}
\rho \frac{S}{N} = && - \nu \int \frac{d^3 p}{h^3} ( f(p) \log f(p) \\ \nonumber && + (1 - f(p)) \log (1 - f(p)) ).
\end{eqnarray}

We may define, for each thermodynamical quantity $X (T,\rho)$, its interaction contribution $X_{int} (T,\rho)$, which is equal to the difference between the quantity calculated for a free FG  at a given temperature $T$ and the quantity itself at the same temperature. For example, for the free energy per nucleon, one may write
\begin{equation}
\label{fintera}
\frac{F_{int} (T,\rho)}{N}=\frac{F_{FG}(T,\rho)}{N}-\frac{F(T,\rho)}{N}.
\end{equation}
The analysis of this quantity indicates how strong are the deviations of the free energy per nucleon from the free FG case. 

We may also define the following difference:
\begin{equation}
X_{th} (T,\rho) = X(T,\rho)-X(T=0, \rho), 
\label{therm}
\end{equation}
which represents the thermal contribution, at a given temperature $T$, of a thermodynamical quantity $X$.

Let us  finally introduce the thermal index $\Gamma_{th}$ as
\begin{equation}
\label{thindex}
\Gamma_{th} (T, \rho)=1+\frac{P_{th}(T,\rho)}{\mathcal{E}_{th}(T,\rho)},
\end{equation}
where the thermal contributions of the pressure 
\begin{equation}
\label{pre}
P(T,\rho)=\rho^2 \frac{\partial}{\partial \rho} \frac{F(T,\rho)}{N}
\end{equation}
and of the energy density $\mathcal{E}=E/V$ are used.

The thermal index is one of the most essential quantities to encompass the thermal effects of dense matter, playing also a crucial role in the neutron-star merger dynamics and ultimately controlling the rapidity of contraction and time delay of a massive star \cite{yasin, bauswein}. However, as
mentioned in the Sec. \ref{intro}, the thermal index is often treated in a very simplified form, and taken to be constant (independent of the temperature and of the density), with values between 1 and 2 \cite{bauswein, constantinou, janka}. As a matter of fact, the thermal index has a dependence on the temperature and the density (except in the case of a free FG, where it is constant and equal to 5/3), being also intimately connected to the density dependence of the effective mass. 
As shown in Ref. \cite{yasin}, for example, larger effective masses lead to lower pressures and a lower thermal index, resulting in a more rapid contraction of the proto-neutron star, which favours the shock evolution to a faster explosion in core-collapse supernovae, due to higher neutrino energies. 

\section{Results}
\label{resu}

Let us focus on PNM, as was done in Ref. \cite{keller}, with the same motivation: in most cases, the scenarios of astrophysical interest where finite-temperature effects are  expected to have an important impact are strongly neutron rich.

\begin{figure}[tp!]
\begin{center}
\includegraphics[width=0.4\textwidth]{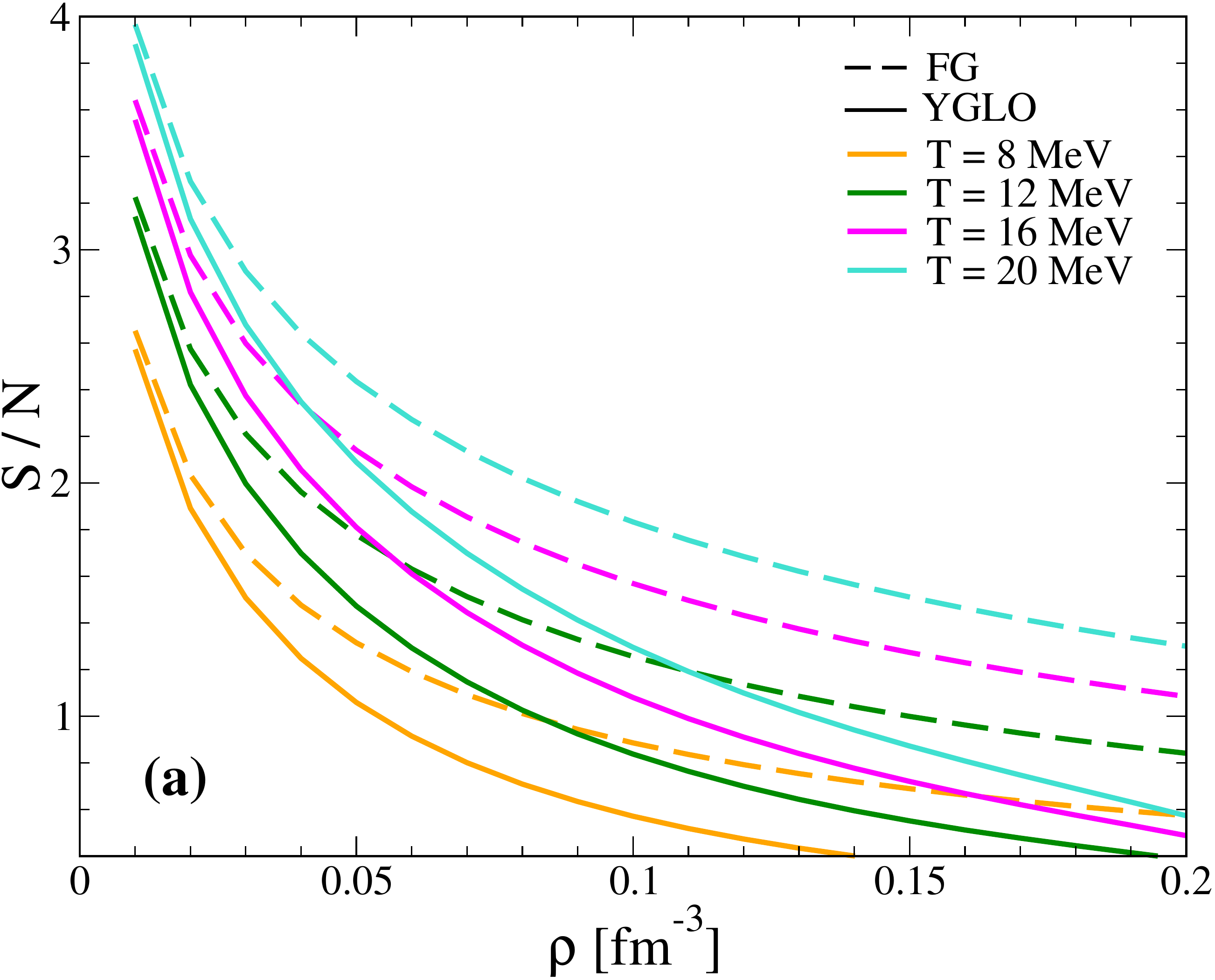} \\ 
\includegraphics[width=0.4\textwidth]{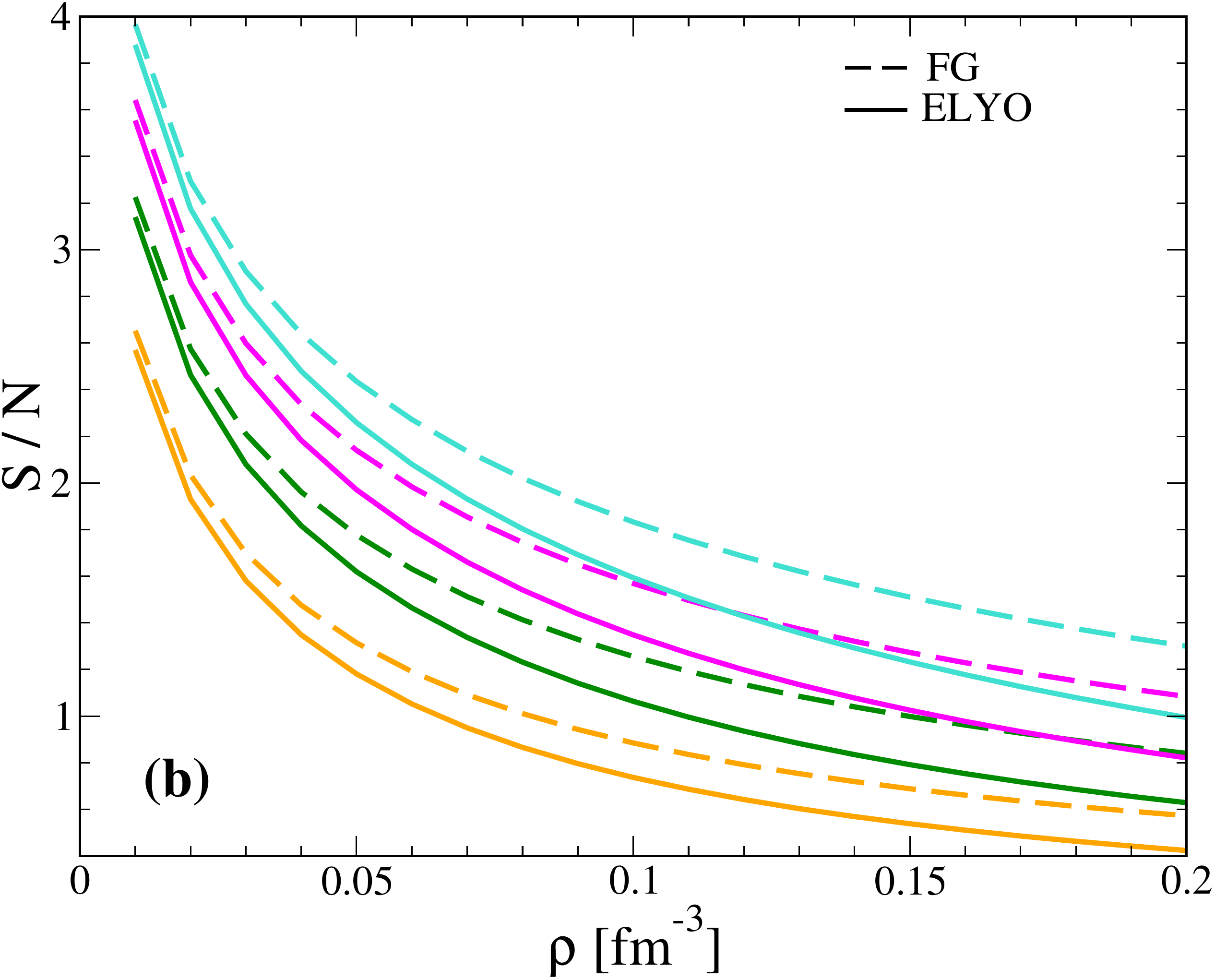} \\
\includegraphics[width=0.4\textwidth]{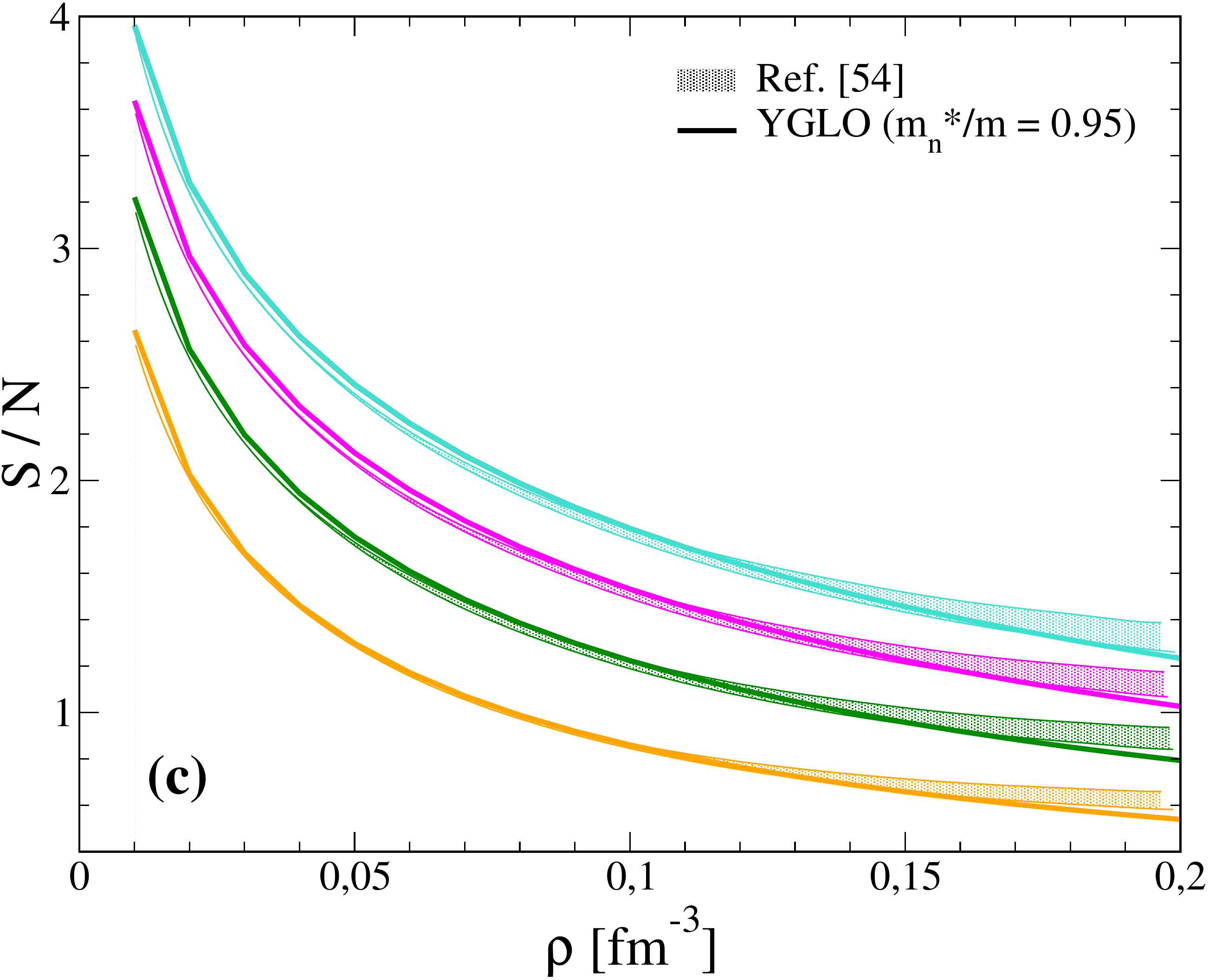} 
\caption{(a) Density dependence of the entropy per nucleon obtained in PNM with YGLO(FP)  and for the Fermi gas (FG); (b) same as in (a) but for ELYO; (c) same as in (a) for the YGLO case, but modifying by hand the splitting parameter $W$ so to produce an effective neutron mass $m_n^* / m$ = 0.95 at $\rho=0.16$ fm$^{-3}$. 
A comparison with bands provided in Ref. \cite{keller} (Fig. 6, right panel) is also shown. 
}
\label{fig:entropy}
\end{center}
\end{figure}

The density dependence of the entropy per nucleon, obtained with YGLO(FP) and ELYO, is displayed in Fig. \ref{fig:entropy} together with the corresponding FG results, for several temperature values. Panel (a) refers to YGLO, panel (b) to ELYO, and panel (c) refers again to the YGLO case obtained by modifying by hand the splitting parameter $W$ so to produce an effective neutron mass $m_n^* / m$ = 0.95 at $\rho_0 = 0.16$ fm$^{-3}$. 
In the latter case, the results are 
compared with the curves of Fig. 6 (right panel) of Ref. \cite{keller}. 
There is no special reason to tune the effective neutron mass to 
this specific value. This choice is 
done only with the aim to consider a wide interval of values for the effective mass and, as upper limit, a neutron effective mass close to the predictions given in Fig. 13 of Ref. \cite{keller}. We notice, by the way, that the adopted value turns out to be even 
closer to 1 than the QMC results of Refs. \cite{bura1, bura2} (see Fig. \ref{fig:effective_mass}). 

One may observe that in Ref. \cite{keller}, at all chosen temperature values and in the selected density interval (the same as in our figure), the entropy per nucleon remains quite close to the free FG result. 
This occurs because the entropy depends on the interaction only through the effective mass. Since the {\it{ab-initio}} effective mass is generally close to the bare value, the results for the entropy come out to be very similar to the free FG curves. 


At low densities, where both EFT-inspired functionals YGLO and ELYO have a neutron effective mass close to the bare mass, our results are also close to the FG ones. Owing to the very low neutron effective mass in the YGLO case at higher densities, strong deviations may be observed there from the FG curves (panel (a)). This behavior is much less pronounced in the ELYO case (panel (b)), for which the neutron effective mass is higher at higher densities.
 The curves get progressively closer to the FG results moving from the top to the bottom of Fig. \ref{fig:entropy} (in the bottom part, panel (c), the parameter $W$ in YGLO is changed by hand so to have a neutron effective mass close to the bare mass). 
As a result, whereas the YGLO functional provides results which are quite different compared to those of Ref. \cite{keller}, especially when densities and temperatures increase, the ELYO functional, with its higher effective mass around saturation, naturally provides values closer to the FG curves, in much better agreement with the corresponding {\it{ab-initio}} results of Ref. \cite{keller}, especially at the lowest temperatures. 

\begin{figure}[t]
\begin{center}
\includegraphics[width=0.4\textwidth]{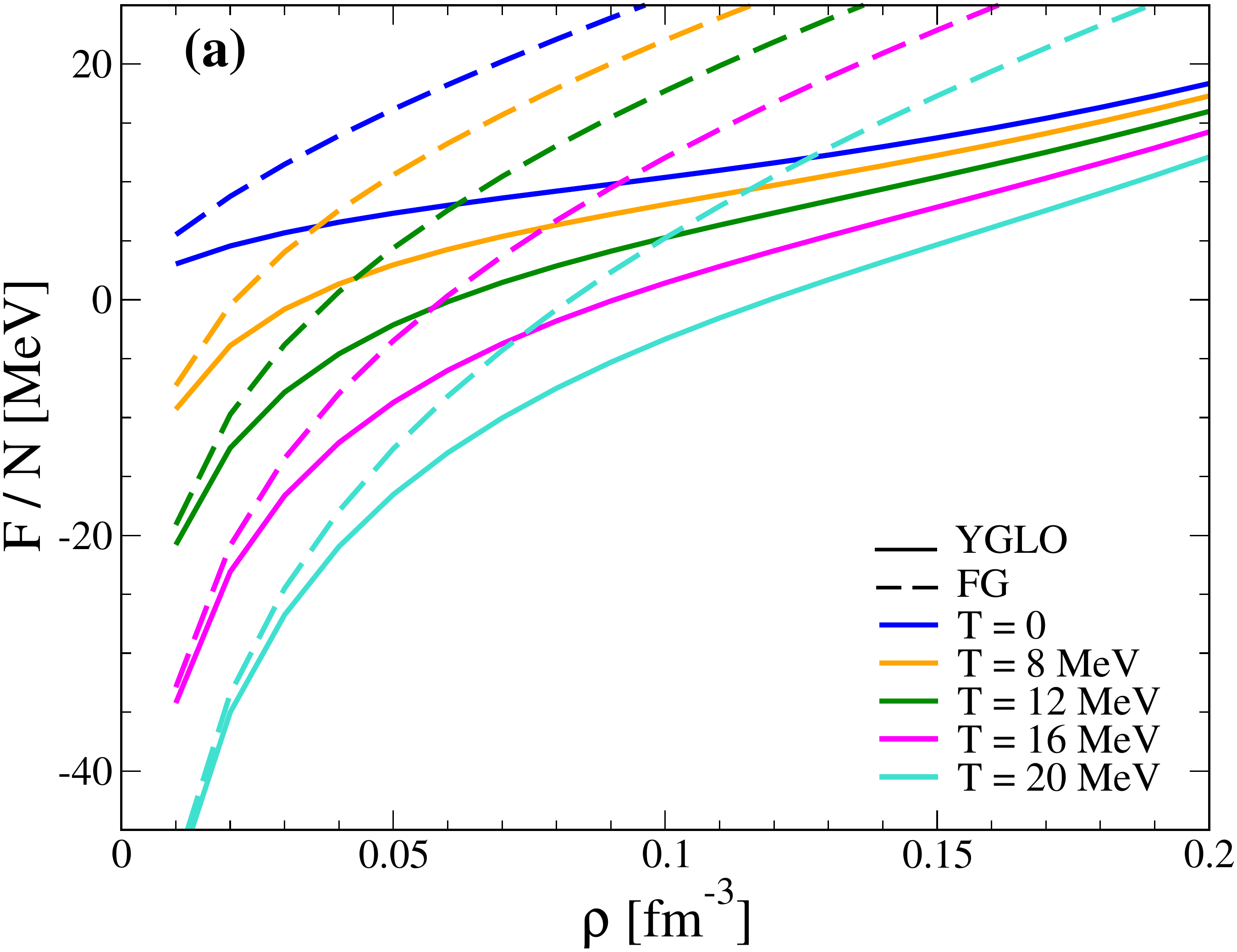} \\ 
\includegraphics[width=0.4\textwidth]{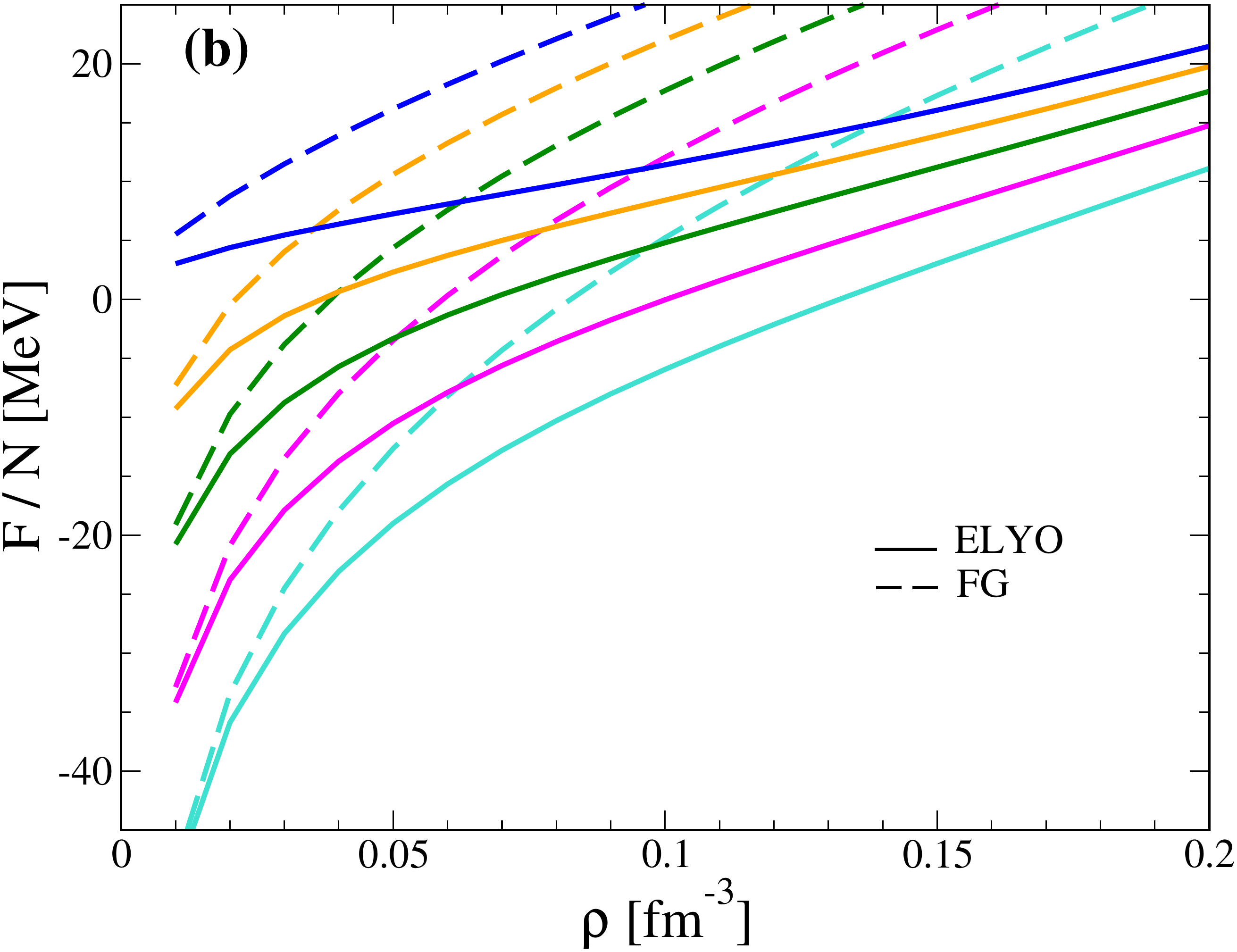} \\
\includegraphics[width=0.4\textwidth]{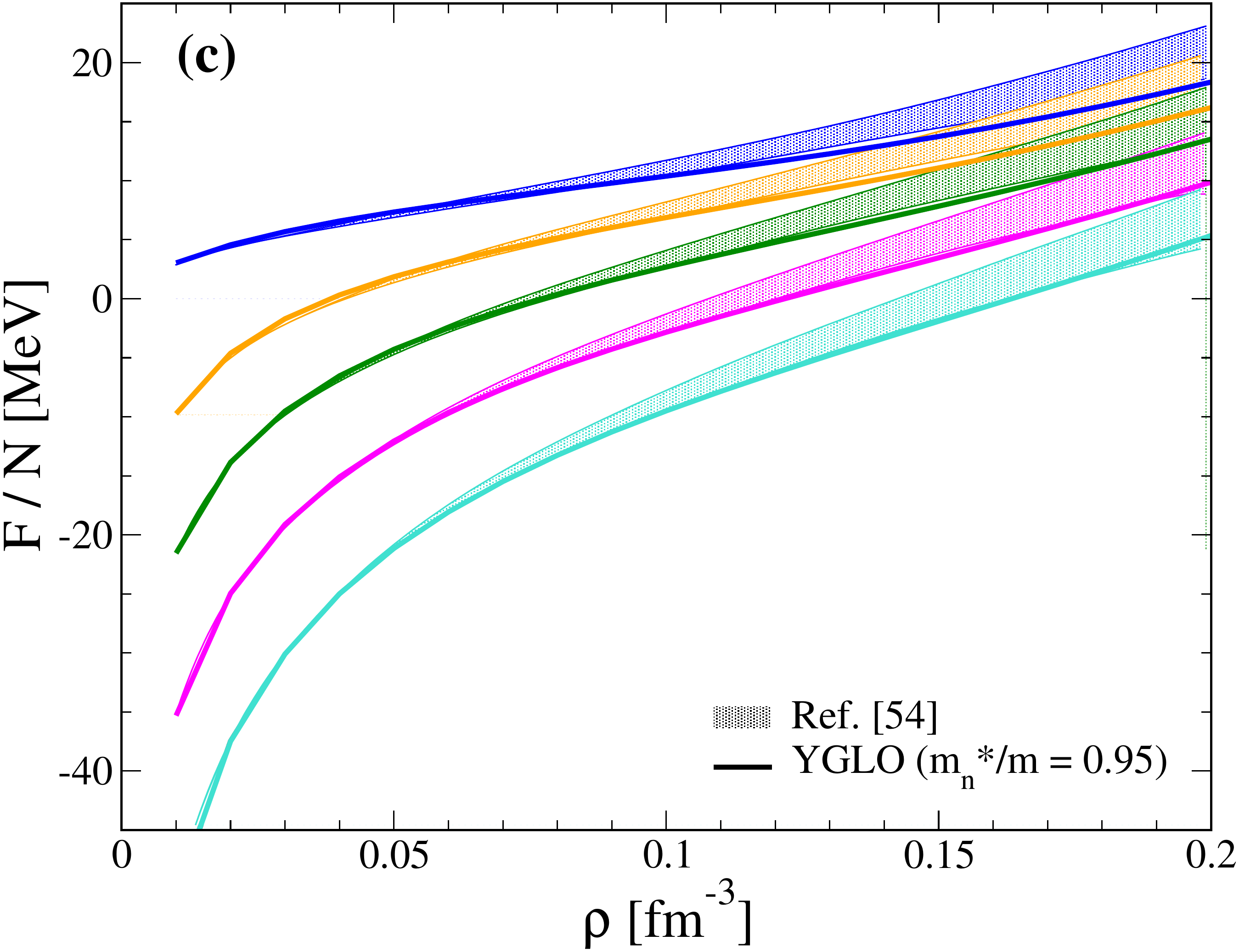} 
\caption{(a) Density dependence of the free energy per nucleon obtained in PNM with YGLO(FP)  and for the Fermi gas (FG); (b) same as in (a) but for ELYO; (c) same as in (a), for the YGLO case, but modifying by hand the splitting parameter $W$ so to produce an effective neutron mass $m_n^* / m$ = 0.95 at $\rho=0.16$ fm$^{-3}$. 
A comparison with bands provided in Ref. \cite{keller} (Fig. 6, left panel) is also shown. 
}\label{fig:free_energy}
\end{center}
\end{figure}

Once the entropy is deduced, taking into account also the energy contribution, one may easily evaluate the free energy per nucleon, whose density dependence  is shown in Fig. \ref{fig:free_energy} for YGLO (a) and ELYO (b). Panel (c) displays, again, the YGLO curves obtained by varying by hand the splitting parameter $W$ so to have a neutron effective mass $m_n^* / m$ = 0.95 at $\rho=0.16$ fm$^{-3}$ 
and compared with the MBPT ones shown in the 
left panel of Fig. 6 in Ref. \cite{keller}. 

One may observe that these results are less strongly affected by the effective mass, as could be easily expected, since the free energy depends also on the other terms of the interaction (not only on the effective-mass contribution, as was the case for the entropy). 
In this case, the full lines strongly depart from the FG ones, owing to the effect of the mean-field potential. The important point is that, also at finite temperatures, the results obtained with the ELYO and YGLO functionals are both rather close to the {\it{ab-initio}} predictions, despite their very different values of effective mass  around saturation. In addition, one may observe that the differences existing at high densities (which are also ascribed to the different stiffnesses of the corresponding EOSs) are strongly reduced at low densities, where both EFT-inspired EDFs are designed in a similar way (both reproduce by construction the LY expansion in the low-density regime).

\begin{figure}[t]
\begin{center}
\includegraphics[width=0.4\textwidth]{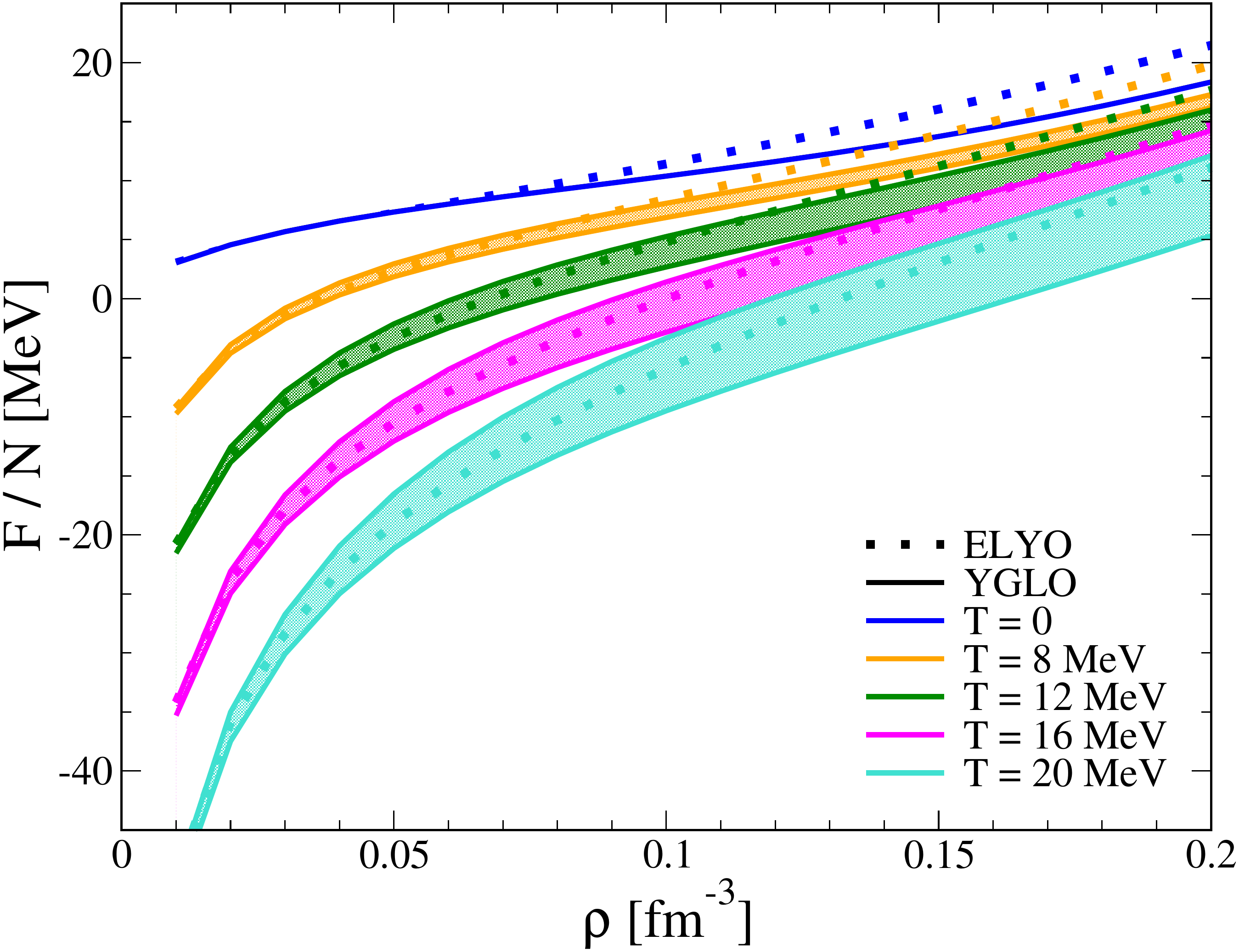} 
\caption{Density dependence of the free energy per nucleon in PNM with the functionals used in Fig. \ref{fig:free_energy}. The bands describe at finite temperature the spread obtained by using the two YGLO(FP) curves of Fig. \ref{fig:free_energy}, panels (a) and (c). Also the 
the FG results are shown. }\label{bands}
\end{center}
\end{figure}

To have a clearer global picture, we resume in Fig. \ref{bands} all the results of panels (a), (b), and (c) of Fig. \ref{fig:free_energy}. The bands in the case of finite temperatures represent the spread in the density dependence of the PNM free energy per nucleon obtained by using the two YGLO(FP) curves of Fig. \ref{fig:free_energy}, panels (a) and (c). 
This is done to mimic the uncertainty that one would produce in the curves by using two functionals related, both, to the same zero-temperature EOS (the YGLO(FP) one in our case) but having different effective mass values. Our window of effective masses is obviously only an illustration and represents what one would expect by moving from a very low effective mass ($m^*_n/m$ equal to 0.500 at saturation) to a value close to the bare mass. 
The ELYO results are shown by dotted lined. At $T=0$, the YGLO and ELYO curves 
practically coincide up to $\rho \approx 0.06$ fm$^{-3}$, as one would expect because they both well describe the low-density regime by construction. 

If the curves at finite temperatures were organized only according to the values of the neutron effective masses, one would expect that the ELYO results should always be located inside the colored bands. However, an ordering according to the values of the effective mass does not hold, 
and there is an interplay between effective-mass effects and the different stiffnesses of the PNM YGLO and ELYO EOSs at densities beyond the low-density regime.     

Considering the colored bands and the ELYO curves, we may observe that the spread of the results in the low-density regime is rather 
small, also at finite temperatures (it increases with the temperature).  
At least when limiting to the density region below 0.05 fm$^{-3}$ and to not too high temperatures (below 10 MeV), 
the deviations are small and the predictions obtained with the YGLO and ELYO functionals are quite similar. Such a result confirms the importance of tailoring functionals to reproduce the correct low-density regime (at zero temperature) by linking them to EFT and {\it{ab-initio}} benchmarks. The found weak spread of the results at low densities and finite temperatures validates the soundness of the procedure adopted for extending the two EFT-inspired functionals 
to account for thermal effects in NM calculations.   Moreover, the spread corresponding to each value of $T$ is not larger than the uncertainties 
exhibited in Ref. \cite{keller} and shown in panel (c) of Fig. \ref{fig:free_energy} (except for the highest temperature $T=20$ MeV), where the spread in the predictions has a different meaning and is related to the use of different chiral interactions. 
This indicates that, in almost all cases, our curves are located inside the {\it{ab-initio}} uncertainties. Such a result implies that 
 both YGLO and ELYO predictions for the free energy per nucleon may be regarded as trustworthy, when compared with  {\it{ab-initio}} results. 

\begin{figure}[tp!]
\begin{center}
\includegraphics[width=0.4\textwidth]{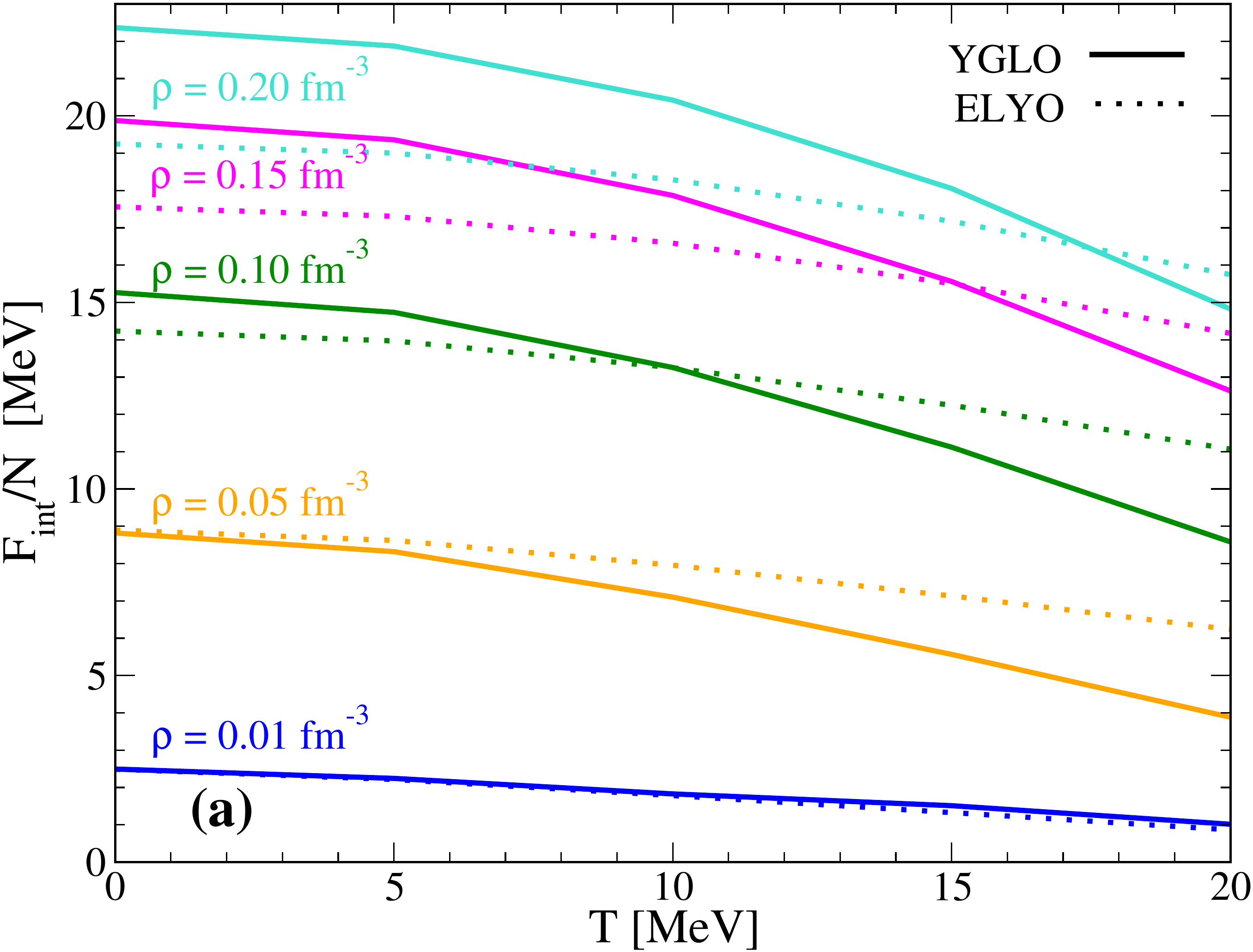} \\ 
\includegraphics[width=0.4\textwidth]{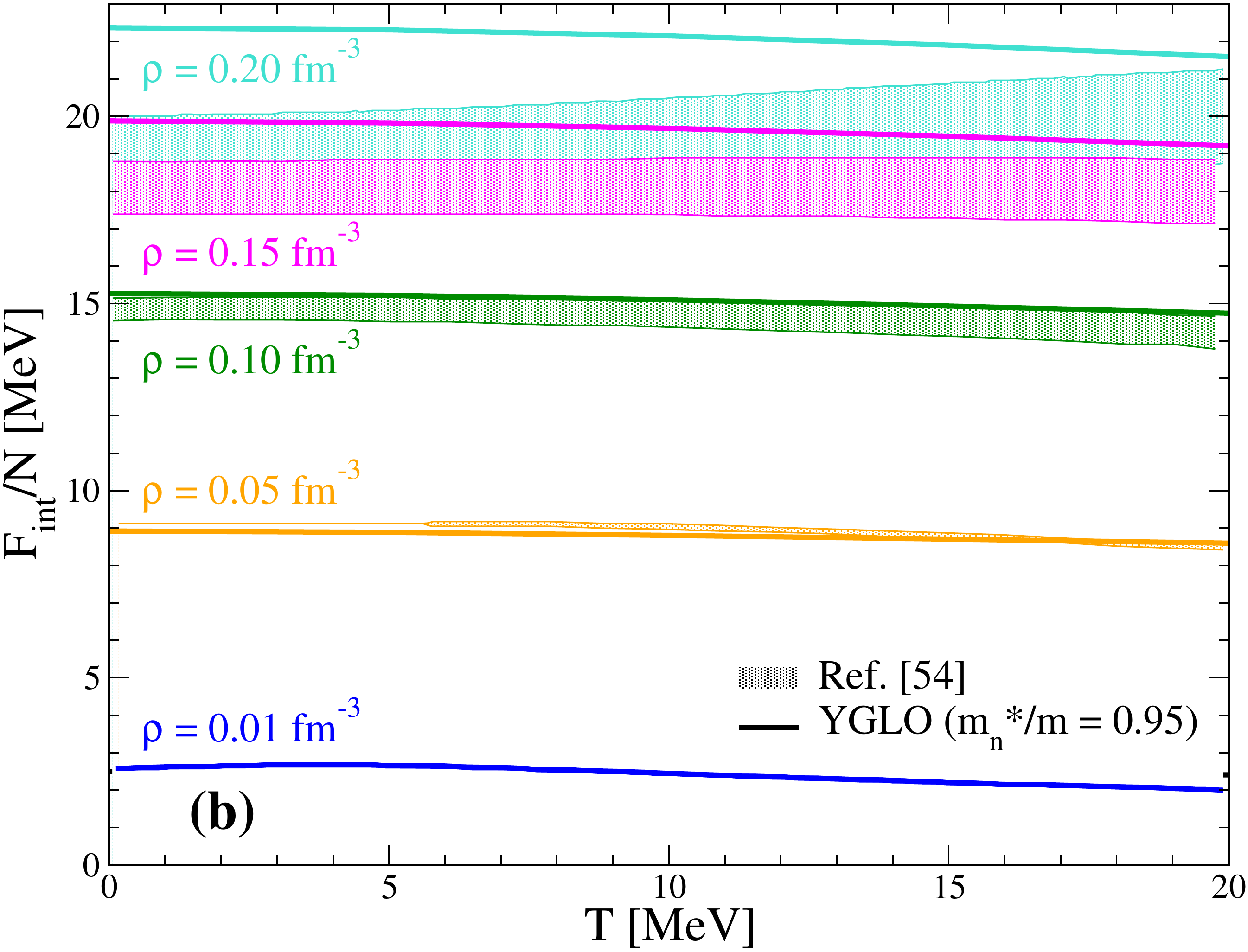} \\
\includegraphics[width=0.4\textwidth]{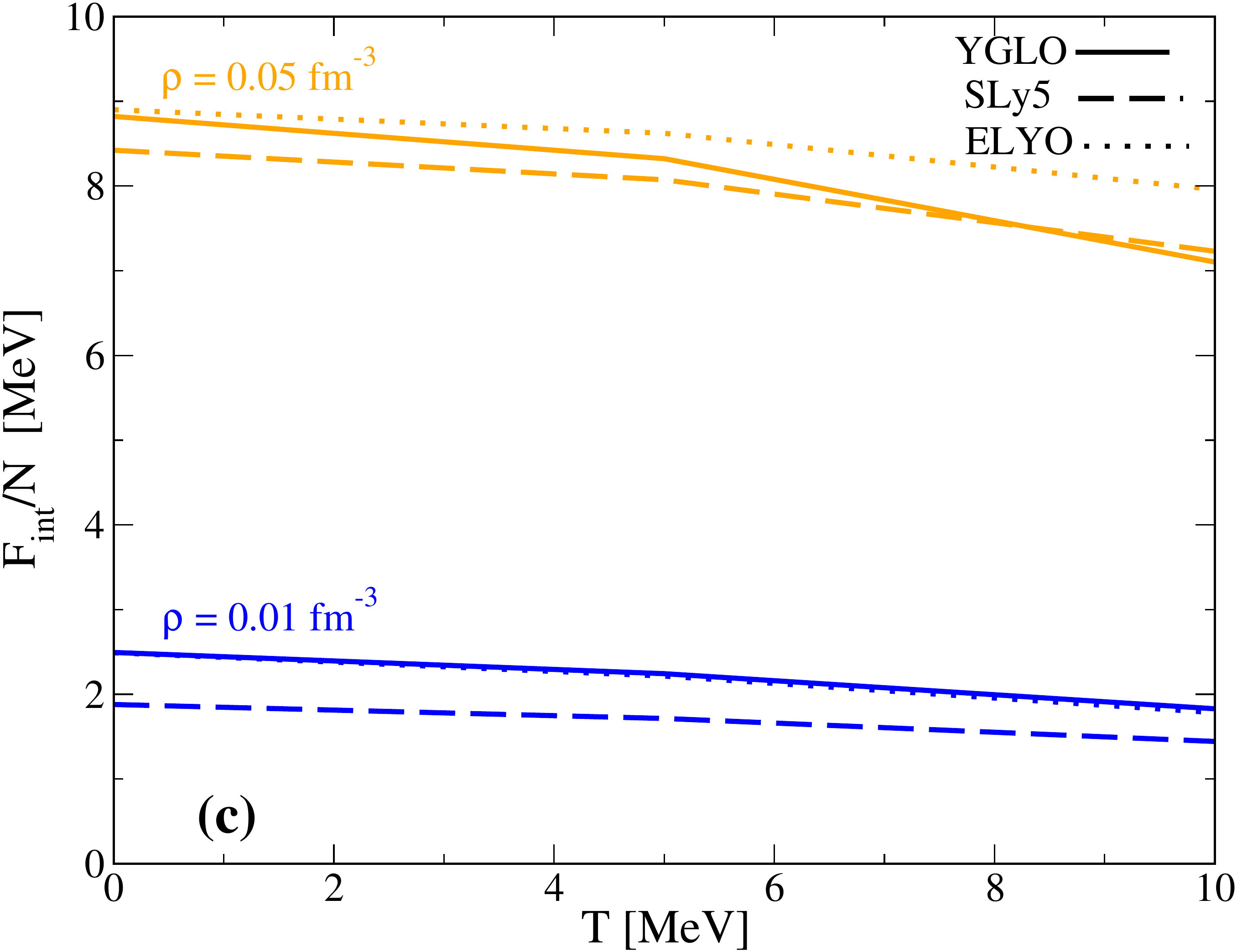} 
\caption{(a) $F_{int}/N$, Eq. (\ref{fintera}), as a function of the temperature, for several selected density values. We show the results obtained with YGLO and ELYO; (b) The same as in (a) but comparing results of Ref. \cite{keller} with the ones obtained by modifying the splitting parameter $W$ of YGLO to have a neutron effective mass $m_n^* / m$ = 0.95 at $
\rho=0.16$ fm$^{-3}$; (c) The same as in (a) but for YGLO, ELYO, and SLy5, at the lowest densities and in a smaller range of temperatures. 
}
\label{fig:free_energy_fg}
\end{center}
\end{figure}

We display in Fig. \ref{fig:free_energy_fg} the quantity $F_{int}/N$, defined in Eq. (\ref{fintera}). The deviations from the FG are plotted, for several values of the density, as a function of the temperature. Panel (a) shows the results for 
YGLO and ELYO. Panel (b) shows the results for the YGLO case with an effective mass increased by hand, as compared with the corresponding ones of Fig. 10 of Ref. \cite{keller}. 
Panel 
(c) 
shows the results for YGLO, ELYO, and the Skyrme functional SLy5, only for the two lowest densities of panel (a), $\rho=0.01$ and 0.05 fm$^{-3}$. 
The curves in Ref. \cite{keller} are quite flat. This indicates that the deviations from the FG, which increase as expected with $\rho$, at a given density remain almost constant as a function of the temperature. Figure \ref{fig:free_energy_fg} (a) shows that YGLO provides a quite flat behavior in the whole interval of temperatures only for the lowest density, $\rho=0.01$ fm$^{-3}$, where the YGLO neutron effective mass is close to the ab-initio predictions. For the other density values,  the curves are quite flat up to temperatures around 5 MeV. Beyond, they show a pronounced decreasing trend, corresponding to the fact that the YGLO effective mass departs from the bare value. When the effective mass is increased by hand, a flat behavior is instead obtained in all cases.

The ELYO curves in panel (a) also show a decreasing trend, as for the YGLO case, but with a sensibly lower slope  also well beyond $T=$ 5 MeV.

It is interesting to notice that, for the two lowest densities in panel (a), YGLO and ELYO tend to the same value of $F_{int}$ when the temperature goes to zero, as should be expected
, since the description of the low-density regime at zero temperature is the same for both functionals by construction. 
As underlined several times, the important peculiarity of the ELYO and YGLO functionals is indeed that, in the limit of zero temperature, they provide the correct behavior for dilute neutron matter. This is not the case, in general, with traditional EDFs such as Skyrme functionals.

We compare in panel 
(c) 
the YGLO and ELYO curves for the two lowest density values, $\rho = 0.01$ and 0.05 fm$^{-3}$, with the corresponding Skyrme SLy5 curves, up to $T=10$ MeV. We observe in the zero-temperature limit the expected result: the SLy5 curves differ from the ELYO and YGLO results (and from the {\it{ab-initio}} predictions of Ref. \cite{keller}). 
On the other side, the YGLO and ELYO functionals reproduce pretty well the {\it{ab-initio}} calculations of Ref. \cite{keller}. We may also see more clearly in this panel of the figure the quite flat behavior up to $T=$ 10 MeV of the YGLO and ELYO curves for  
$\rho = 0.01$ fm$^{-3}$ and of the ELYO curve for $\rho = $ 0.05 fm$^{-3}$.

\begin{figure}[h]
\centering
\includegraphics[width=0.4\textwidth]{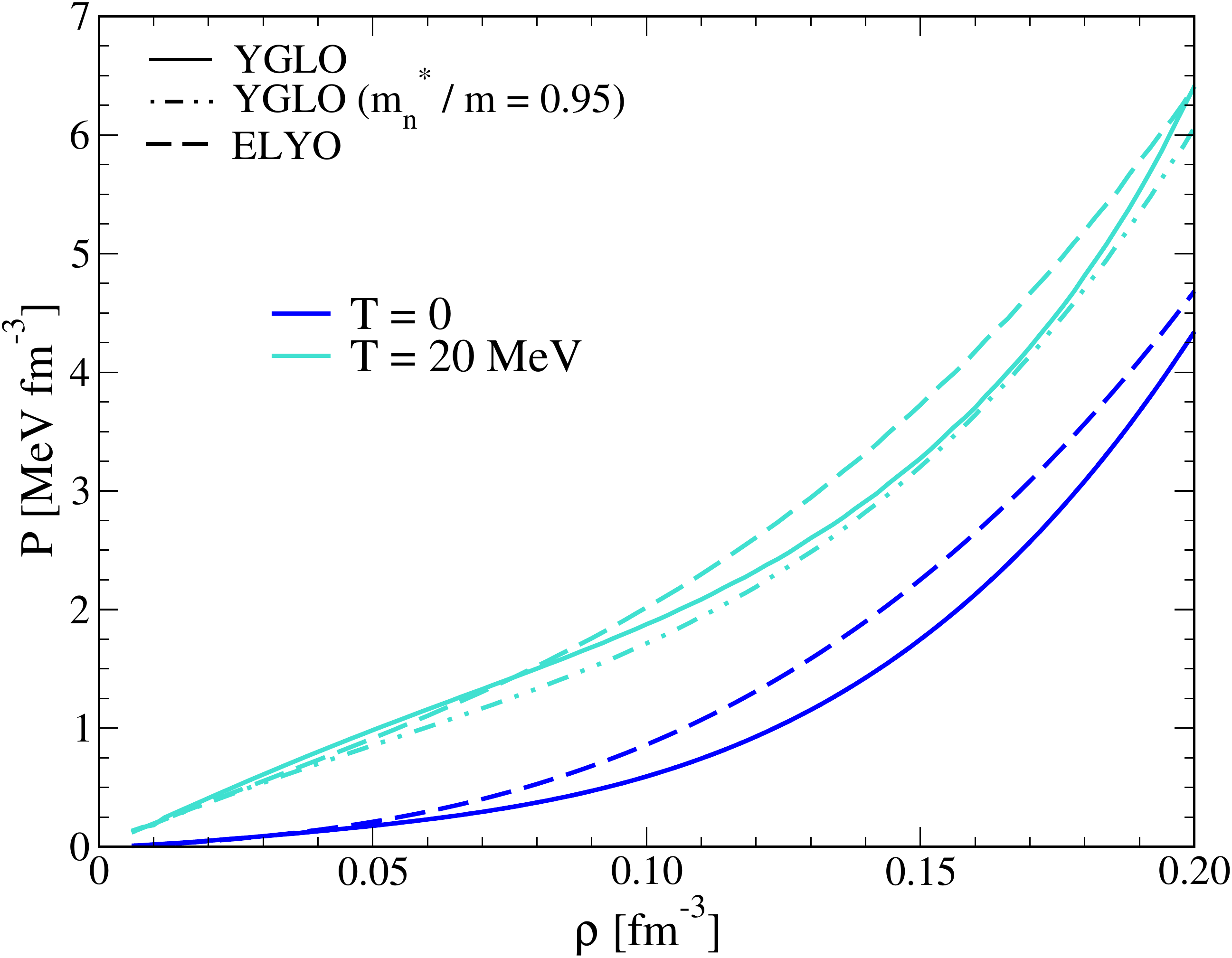}
\caption{Pressure as a function of the density for $T=$ 0 and $T=$ 20 MeV computed with YGLO, ELYO, and the YGLO case where the splitting parameter $W$ is modified to have a neutron effective mass $m_n^* / m$ = 0.95 at $
\rho=0.16$ fm$^{-3}$.} 
\label{fig:pressure}
\end{figure}

Figure \ref{fig:pressure} represents the pressure computed at zero temperature and at $T=$ 20 MeV. This figure can be compared with the two panels of Fig. 5 and with the right panel of Fig. 9 of Ref. \cite{keller}. 
The modification of the effective mass in the YGLO case  has a very weak effect on the curves. The results are (obviously) strongly driven by the different stiffnesses of the YGLO and ELYO EOSs. Globally, our curves follow the corresponding {\it{ab-initio}} predictions, within the uncertainty band (related to the choice of the chiral interaction) which characterizes those results. 

\begin{figure}[t]
\begin{center}
\includegraphics[width=0.4\textwidth]{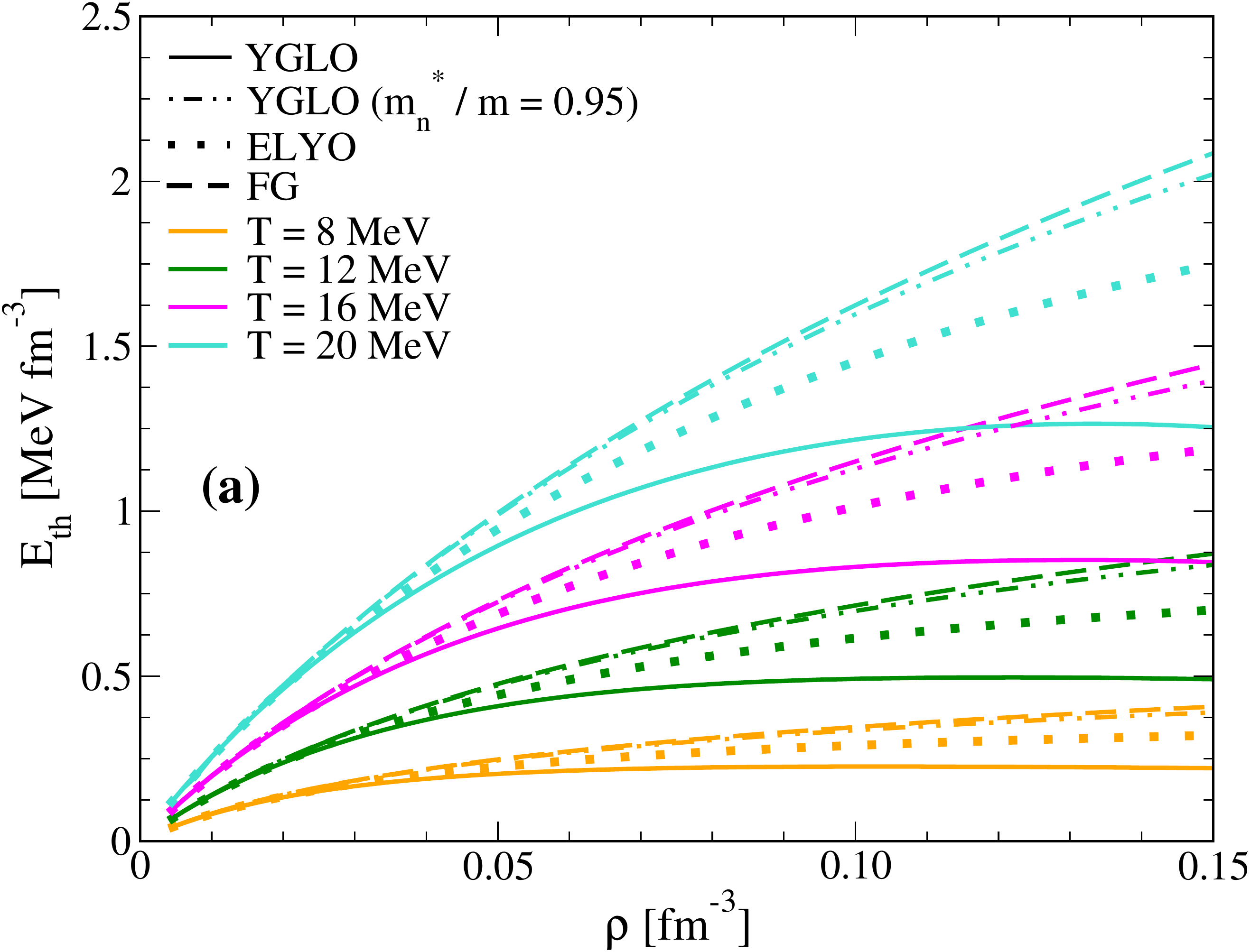} \\ 
\includegraphics[width=0.4\textwidth]{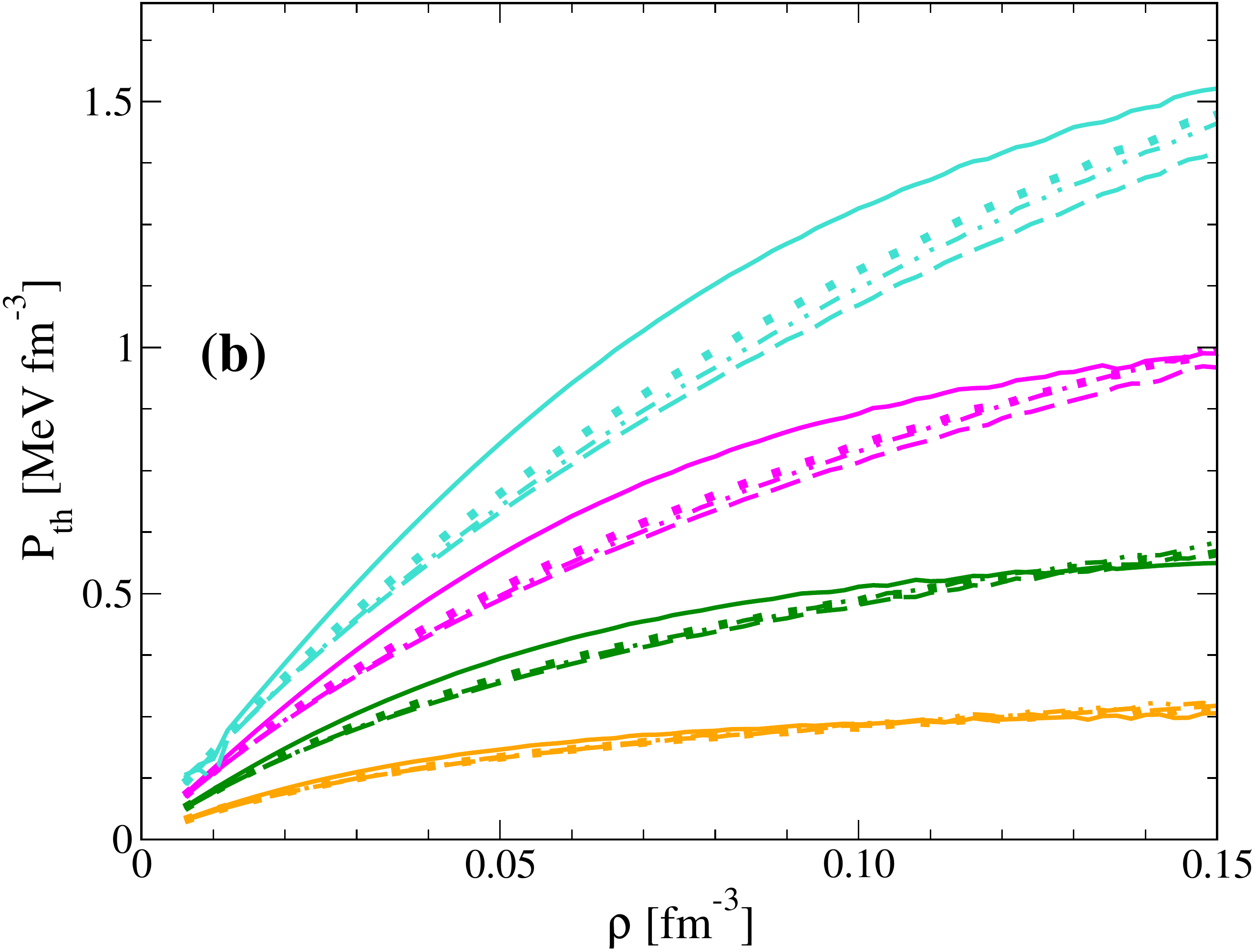} \\
\includegraphics[width=0.4\textwidth]{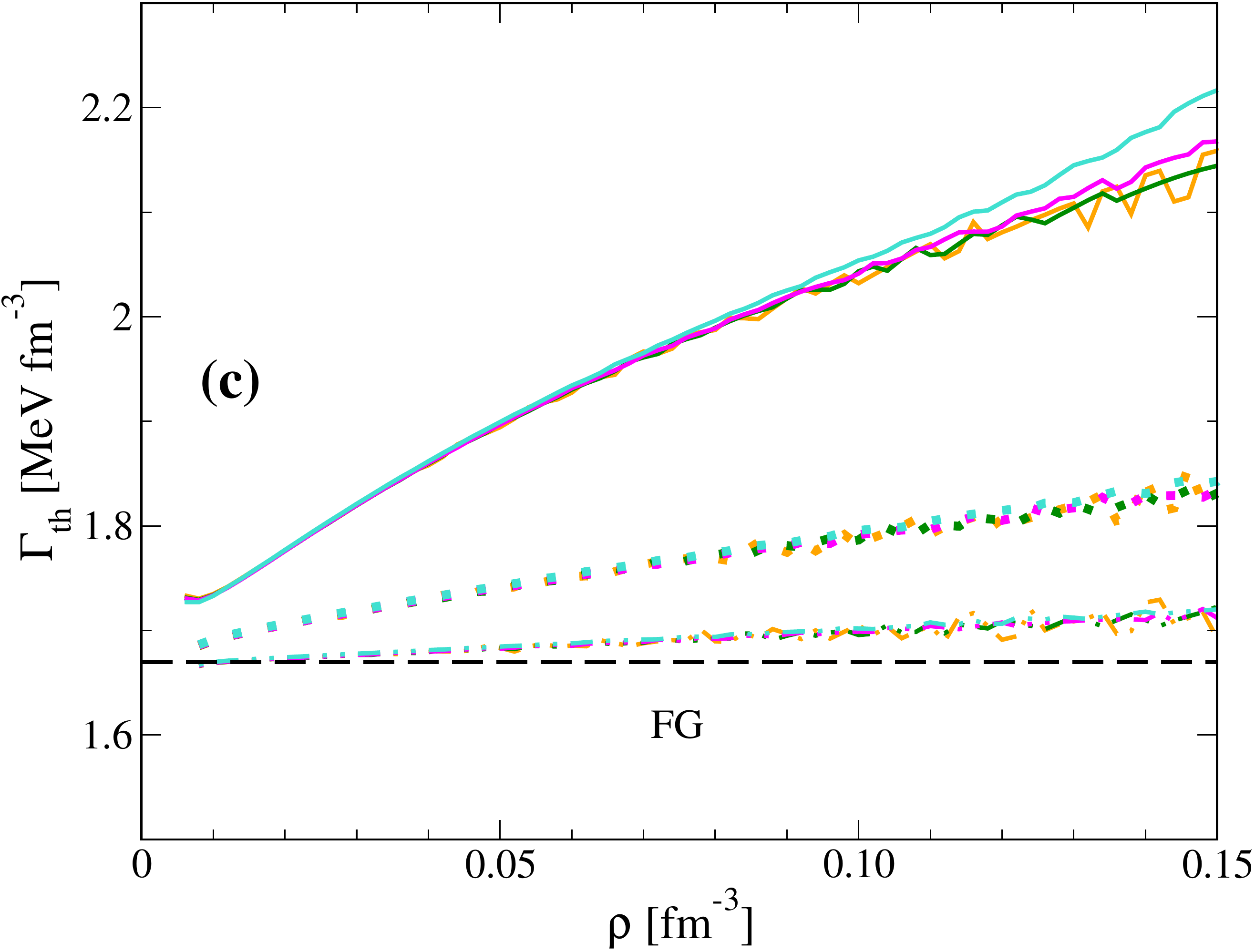} 
\caption{Thermal energy (a), thermal pressure (b), and thermal index (c) as a function of the density for different temperatures. YGLO and ELYO results are shown, together with the YGLO curves obtained with the modified splitting parameter.}\label{fig:gamma}
\end{center}
\end{figure}

We finally show in the three panels of Fig. \ref{fig:gamma} the thermal energy $E_{th}$ (a), the thermal pressure $P_{th}$ (b), and the thermal index $\Gamma_{th}$ (c) obtained at various temperatures in the YGLO and ELYO cases, 
as well as in the YGLO case with a modified splitting parameter.
The curves corresponding to $T=$ 20 MeV 
can be compared with the three panels of Fig. 11 of Ref. \cite{keller}, where the same quantities at this value of temperature are plotted. 

The {\it{ab-initio}} results at $T=$ 20 MeV show for the thermal energy a very weak dependence on the interaction.  On the other side, the thermal pressure and the thermal index differ from the free FG curves.
 
In our case, due to the very low effective mass around saturation, the YGLO curves for the thermal energies are quite different from the corresponding free FG curves. Such differences become more pronounced when densities and temperatures increase, leading to results that definitely disagree from the {\it{ab-initio}} ones at $T=$ 20 MeV. When the effective mass is increased by hand in the YGLO case, the agreement with the {\it{ab-initio}} results at $T=20$ MeV is sensibly improved, as expected. The ELYO case is intermediate owing to its intermediate neutron effective mass. One may observe that the agreement with the $T=$ 20 MeV {\it{ab-initio}} results remains quite reasonable up to densities around 0.1 fm$^{-3}$. 

For the thermal pressure we may see that, at $T=20$ MeV, the YGLO curve is, at all densities shown in the figure, above the corresponding free FG result, differently from what is observed for the 
{\it{ab-initio}} curve. The ELYO and the modified YGLO curves, on the other side, follow more closely the free FG results, remaining in a quite reasonable agreement with the {\it{ab-initio}} trend up to densities between 0.1 and 0.15 fm$^{-3}$. 
Such a result, which is ascribable to the unsatisfactory reproduction of the $ab-initio$ results for the effective mass with the YGLO functional, confirms once again the crucial importance of this ingredient to properly account for the thermal effects of dense matter. 

The thermal index is shown in the bottom panel and displays a weak temperature dependence (as one can also see in Fig. 12 of Ref. \cite{keller}). 
One may observe in this panel the preminent role played by the density behavior of the effective mass, whose relevance is larger than the stiffness of the EOS. While for density values below 0.02 fm$^{-3}$, the value of the thermal index remain rather well constrained below 1.8 in all cases, at higher densities the density behavior of the effective mass strongly affects this quantity. As a result, an important disagreement is observed between the YGLO curve and the {\it{ab-initio}} results of Ref. \cite{keller}. The trend is strongly modified when the YGLO splitting parameter is changed by hand whereas, again, the ELYO case is intermediate. The curve is much flatter than for YGLO and one may observe that, at least up to densities around 0.1 fm$^{-3}$, the value of the thermal index remains lower than 1.8, in reasonable agreement with the {\it{ab-initio}} values (inside their uncertainty band). 

\section{Conclusions}
\label{conclu}

The recently introduced EFT-inspired functionals ELYO and YGLO are employed for the first time in this work to describe finite-tempetature properties of PNM.  Our objective is to  verify whether these functionals, which are both designed to correcly describe the very low-density regime of neutron matter, allow us to correctly incorporate finite-temperature effects. 
For this, we compare the results obtained for some selected thermodynamical quantities with the corresponding {\it{ab-initio}} ones obtained in Ref. \cite{keller} using the MBPT approach. We compare in particular  the entropy per nucleon, the free energy per nucleon, the pressure, the thermal energy, the thermal pressure, and the thermal index. 

We conclude that, globally, the EFT-inspired ELYO and YGLO functionals provide both quite reliable results at least at low densities and not too high temperatures, 
with the exception of the thermal energy, the thermal pressure, and the thermal index, where YGLO predictions strongly deviate from {\it{ab-initio}} results. 

We have discussed in particular the effects related to the effective mass and we  may conclude that, at the present stage, the ELYO functional, having a higher neutron effective mass around saturation (in better agreement with {\it{ab-initio}} values) allows us to describe finite-temperature properties of PNM more satisfactorily than the YGLO functional (up to higher densities and temperatures). In particular, the ELYO thermal energy, thermal pressure, and thermal index come out to be closer to {\it{ab-initio}} predictions.   

As mentioned in Ref. \cite{yglonuclei}, possible directions for increasing the YGLO effective mass are envisaged, by include also  $p$-wave contributions, while still benchmarking on ab-initio results for both EOS and neutron drop energies. Further constraints would be moreover advisable in order to possibly disentangle and recover also the thermal effects connected to the temperature evolution of the pairing correlations, whose importance within this context has been recently addressed (see Refs. \cite{shelley, burrello14}).

The possibility of using these EFT-inspired functionals for describing finite-temperature properties of neutron matter opens new and promising horizons for computations and modeling in scenarios of astrophysical interest. The description of the low-density regime of matter has indeed revealed itself to be extremely important, to properly evaluate a broad spectrum of observables of astrophysical relevance, such as the heat capacity, the thermal conductivity, the neutrino emission, as well as for a precise determination of the crust-core transition point and the related crustal moment of inertia \cite{fortin, burrello16, oertel20, thomas, boulet2, bura3, bura4, bura5}. 

\acknowledgements{The authors acknowledge fruitful discussions with J\'er\'emy Bonnard. M.G. acknowledges funding from the IN2P3-CNRS BRIDGE-EDF project. S. B. acknowledges support from the Alexander von Humboldt foundation.}


\begin{thebibliography}{99}
\bibitem{bender} M. Bender, P. H. Heenen, and P. G. Reinhard, Rev. Mod. Phys. {\bf 75}, 121 (2003).
\bibitem{colo} G. Col\`o, Advances in Physics: X {\bf 5}, 1740061 (2020), https://doi.org/10.1080/23746149.2020.1740061.
\bibitem{schunck} N. Schunck, ed. {\it{Energy Density Functional Methods for Atomic Nuclei}}, 2053-2563 (IOP Publishing, 2019).
\bibitem{nakatsukasa} T. Nakatsukasa, K. Matsuyanagi, M. Matsuo, and K. Yabana, Rev. Mod. Phys. {\bf 88}, 045004 (2016).
\bibitem{stone} J. Stone and P.-G. Reinhard, Progress in Particle and Nuclear Physics {\bf 58}, 587–657 (2007).
\bibitem{negele} J. W. Negele, D. Vautherin, Nucl. Phys. A {\bf 207}, 298 (1973).
\bibitem{fantina} A. F. Fantina, J. L. Zdunik, N. Chamel, J. M. Pearson, P. Haensel, S. Goriely, A\&A {\bf 620}, A105 (2018).
\bibitem{potekhin} A. Y. Potekhin, A. F. Fantina, N. Chamel, J. M. Pearson, S. Goriely, A\&A {\bf 560}, A48 (2013).
\bibitem{pearson1} J. M. Pearson, S. Goriely, N. Chamel, Phys. Rev. C {\bf 83}, 065810 (2011).
\bibitem{pearson2} J. M. Pearson, N. Chamel, S. Goriely, and C. Ducoin, Phys. Rev. C {\bf 85}, 065803 (2012).
\bibitem{burrello15} S. Burrello, F. Gulminelli, F. Aymard, M. Colonna and Ad. R. Raduta, Phys. Rev. C {\bf 92}, 055804 (2015).
\bibitem{yangpc} C. J. Yang, M. Grasso, and D. Lacroix, Phys. Rev. C {\bf 96}, 034318 (2017).  
\bibitem{burrello} S. Burrello, M. Grasso, and C. J. Yang, Phys. Lett. B {\bf 811}, 135938 (2020).  
\bibitem{fur1} R. J. Furnstahl, in {\{Renormalization Group and Effective Field Theory Approaches to Many-Body Systems}, edited by A. Schwenk and J. Polonyi, Lecture Notes in Physics {\bf 852} (Springer, Berlin, 2012), Chap. 3.
\bibitem{fur2} R. J. Furnstahl, Eur. Phys. J. A {\bf 56}, 85 (2020).
\bibitem{grasso2019} M. Grasso, Prog. Part. Nucl. Phys. {\bf 106}, 256 (2019). 
\bibitem{yglo2016} C. J. Yang, M. Grasso, D. Lacroix, Phys. Rev. C {\bf 94}, 031301(R) (2016). 
\bibitem{elyo2017} M. Grasso, D. Lacroix, and C. J. Yang, Phys. Rev. C {\bf 95}, 054327 (2017).
\bibitem{drop1} J. Bonnard, M. Grasso, and D. Lacroix, Phys. Rev. C {\bf 98}, 034319 (2018); Phys. Rev. C {\bf 103}, 039901(E) (2021).
\bibitem{drop2} J. Bonnard, M. Grasso, and D. Lacroix, Phys. Rev. C {\bf 101}, 064319 (2020).
\bibitem{yglonuclei} S. Burrello, J. Bonnard, and M. Grasso, Phys. Rev. C {\bf 103}, 064317 (2021).
\bibitem{LY1} T. D. Lee and C. N. Yang, Phys. Rev. {\bf 105}, 1119 (1957). 
\bibitem{LY2} K. Huang and C. N. Yang, Phys. Rev. {\bf 105}, 767 (1957). 
\bibitem{LY3} V. N. Efimov, M. Ya. Amusia, Sov. Phys. JETP {\bf 20}, 388 (1965).
\bibitem{LY4} G. A. Baker, Rev. Mod. Phys. {\bf 43}, 479 (1971).
\bibitem{LY5} R. F. Bishop, Ann. Phys. {\bf 77}, 106 (1973).
\bibitem{LY6} M. Ya. Amusia and V. N. Efimov, Ann. Phys. (NY) {\bf 47}, 377 (1968).
\bibitem{arianna} A. Carbone and A. Schwenk, Phys. Rev. C {\bf 100}, 025805 (2019).
\bibitem{oertel} M. Oertel, M. Hempel, T. Kl\"{a}hn, and S. Typel, Rev. Mod. Phys. {\bf 89}, 015007 (2017).
\bibitem{schneider} A. S. Schneider, L. F. Roberts, C. D. Ott, and E. O'Connor, Phys. Rev. C {\bf 100}, 055802 (2019).
\bibitem{yasin} H. Yasin, S. Sch\"{a}fer, A. Arcones, and A. Schwenk, Phys. Rev. Lett. {\bf 124}, 092701 (2020).
\bibitem{burgio} G. F. Burgio and  A. F. Fantina, Astrophys. Space  Sci. Libr. {\bf 457}, 255 (2018).
\bibitem{bauswein} A. Bauswein, H.-T. Janka, and R. Oechslin, Phys. Rev. D {\bf 82}, 084043 (2010).
\bibitem{constantinou} C. Constantinou, B. Muccioli, M. Prakash, and J. M. Lattimer, Phys. Rev. C {\bf 92}, 025801 (2015).
\bibitem{arianna2} A. Carbone and A. Schwenk, Phys. Rev. C {\bf 100}, 025805 (2019).
\bibitem{shibata1} M. Shibata, Phys. Rev. Lett. {\bf 94}, 201101 (2005).
\bibitem{shibata2} M. Shibata, K. Taniguchi, and K. Uryu, Phys. Rev. D {\bf 71}, 084021 (2005).
\bibitem{shibata3} M. Shibata and K. Taniguchi, Phys. Rev. D {\bf 73}, 064027 (2006).
\bibitem{kiuchi} K. Kiuchi, Y. Sekiguchi, M. Shibata, and K. Taniguchi, Phys. Rev. D {\bf 80}, 064037 (2009).
\bibitem{janka} H.-T. Janka, T. Zwerger, and R. Monchmeyer, Astron. Astrophys. {\bf 268}, 360 (1993).
\bibitem{du} X. Du, A. W. Steiner,  and J. W. Holt, Phys. Rev. C {\bf 99}, 025803 (2019).
\bibitem{raithel} C. A. Raithel, F. \"{O}zel,  and D. Psaltis, Astrophys. J {\bf 875}, 12 (2019).
\bibitem{FP} B. Friedman and V. Pandharipande, Nucl. Phys. A {\bf 361}, 502 (1981).
\bibitem{zuo} W. Zuo, Z. Li, and U. Lombardo, Nucl. Phys. A {\bf 745}, 34 (2004).
\bibitem{tolos} L. Tolos, B. Friman, and A. Schwenk, Nucl. Phys. A {\bf 806}, 105 (2008). 
\bibitem{fiorilla} S. Fiorilla, N. Kaiser, and W. Weise, Nucl. Phys. A {\bf 880}, 65 (2012). 
\bibitem{carbone1} A. Carbone, A. Polls, and A. Rios, Phys. Rev. C {\bf 88}, 044302 (2013). 
\bibitem{welle1} C. Wellenhofer, J.W. Holt, N. Kaiser, and W. Weise, Phys. Rev. C {\bf 89} 064009 (2014). 
\bibitem{welle2} C. Wellenhofer, J.W. Holt, and N. Kaiser, Phys. Rev. C {\bf 92}, 015801 (2015). 
\bibitem{welle3} C. Wellenhofer, J.W. Holt, and N. Kaiser, Phys. Rev. C {\bf 93}, 055802 (2016). 
\bibitem{carbone2} A. Carbone, A. Polls, and A. Rios, Phys. Rev. C {\bf 98}, 025804 (2018). 
\bibitem{rios} A. Rios, Front. Phys. {\bf 8}, 387 (2020).
\bibitem{lu} B. N. Lu, N. Li, S. Elhatisari, D. Lee, J. E. Drut, T. A. L\"{a}hde, E. Epelbaum, U.-G. Meissner, Phys. Rev. Lett. {\bf 125}, 192502 (2020). 
\bibitem{keller} J. Keller, C. Wellenhofer, K.  Hebeler, and  A.  Schwenk, Phys. Rev. C {\bf 103}, 055806 (2021).
\bibitem{sk1} T. H. R. Skyrme, Phil. Mag. 1 (1956) 1043.
\bibitem{sk2} T. H. R. Skyrme, Nucl. Phys. {\bf 9}, 615 (1959).
\bibitem{vau} D. Vautherin and D. M. Brink, Phys. Rev. C {\bf 5}, 626 (1972).
\bibitem{zheng2} H. Zheng, S. Burrello, M. Colonna, and V. Baran, Phys. Lett. B {\bf 769}, 424-429 (2017).
\bibitem{gezerlis} A. Gezerlis and J. Carlson, Phys. Rev. C {\bf 81}, 025803 (2010).
\bibitem{akmal} A. Akmal, V. R. Pandharipande and D. G. Ravenhall, Phys. Rev. C {\bf 58}, 1804 (1998).
\bibitem{cha} E. Chabanat, P. Bonche, P. Haensel, J. Meyer, R. Schaeffer, Nucl. Phys. A {\bf 627}, 710 (1997); {\bf 635}, 231 (1998); {\bf 643}, 441 (1998).
\bibitem{dri} C. Drischler, V. Som\`a, and A. Schwenk, Phys. Rev. C {\bf 89}, 025806 (2014).
\bibitem{sch} A. Schwenk, B. Friman, and G.E. Brown, Nucl. Phys. A {\bf 713}, 191 (2003).
\bibitem{wam} J. Wambach, T. L. Ainsworth, and D. Pines, Nucl. Phys. A {\bf 555}, 128 (1993). 
\bibitem{bura1} M. Buraczynski, N. Ismail, and A. Gezerlis, Phys. Rev. Lett. {\bf 122}, 152701 (2019).
\bibitem{bura2} M. Buraczynski, N. Ismail, and A. Gezerlis, Eur. Phys. J. A {\bf 56}, 112 (2020).
\bibitem{boulet} D. Lacroix, A. Boulet, M. Grasso, and C.-J. Yang, Phys. Rev. C {\bf 95}, 054306 (2017).
\bibitem{shelley} M. Shelley and A. Pastore, Phys. Rev. C {\bf 103}, 035807 (2021).
\bibitem{burrello14} S.  Burrello, M. Colonna, and F. Matera, Phys. Rev. C {\bf 89}, 057604 (2014).
\bibitem{fortin} M. Fortin, F. Grill, J. Margueron, Dany Page, and N. Sandulescu, Phys. Rev. C {\bf 82}, 065804 (2010).
\bibitem{burrello16} S. Burrello, M. Colonna, and F. Matera, Phys. Rev. C {\bf 94}, 012801(R) (2016).
\bibitem{oertel20} M. Oertel, A. Pascal, M. Mancini, and J. Novak, Phys. Rev. C {\bf 102}, 035802 (2020).
\bibitem{thomas} T. Carreau, F. Gulminelli, and J. Margueron, Phys. Rev. C {\bf 100}, 055803 (2019).
\bibitem{boulet2} A. Boulet and D. Lacroix, Phys. Rev. C {\bf 97}, 014301 (2018).
\bibitem{bura3} M. Buraczynski and A. Gezerlis, Phys. Rev. Lett. {\bf 116}, 152501 (2016).
\bibitem{bura4} M. Buraczynski and A. Gezerlis, Phys. Rev. C {\bf 95}, 044309 (2017).
\bibitem{bura5} M. Buraczynski, S. Martinello, and A. Gezerlis, Phys. Lett. B {\bf 818}, 136347 (2021).

\end{thebibliography}
\end{document}